\documentclass[10pt]{elsarticle}
\usepackage{amsmath, amsfonts, amsthm, amssymb, verbatim,dsfont}
\usepackage{graphicx}
\usepackage[mathcal]{euscript}
\usepackage[applemac]{inputenc}
\usepackage{dsfont} 
\usepackage{amssymb,mathrsfs}
\usepackage[usenames]{color}
\usepackage{hyperref} 

\usepackage[T1]{fontenc}%

\theoremstyle{plain}

\theoremstyle{definition}

\theoremstyle{plain}
        
\numberwithin{equation}{section}

\usepackage{amsmath}
\usepackage{verbatim}
\newcommand \be           {\begin{equation}}
\newcommand \ee            {\end{equation}}

\newcommand \RR           {\mathbb{R}}

\newcommand \Pbold           {\mathbf{P}} 
\newcommand \PP \Pbold

\newcommand \del           \partial
\newcommand \eps            \epsilon

\DeclareMathOperator    \dive {\nabla\cdot}
\newcommand{\uref}{u_{\mathrm{ref}}}
\newcommand{\cref}{c_{\mathrm{ref}}}
\newcommand{\nn}{{\mathbf{n}}}
\newcommand{\cm}{{\mathrm{cm}}}
\newcommand{\seg}{{\mathrm{s}}}
\def\XXint#1#2#3{{\setbox0=\hbox{$#1{#2#3}{\int}$}
\vcenter{\hbox{$#2#3$}}\kern-.5\wd0}}

\let\oldmarginpar\marginpar
\renewcommand\marginpar[1]{\-\oldmarginpar[\raggedleft\footnotesize #1]%
{\raggedright\footnotesize #1}}
\def\build#1_#2^#3{\mathrel{
\mathop{\kern 0pt#1}\limits_{#2}^{#3}}}
\allowdisplaybreaks[1]

\begin{document}

\begin{frontmatter}

\title{Modeling ant foraging: a chemotaxis approach with pheromones and trail formation}



\author{Paulo Amorim}
\address{Instituto de Matem\'atica, Universidade Federal do Rio de Janeiro,
Av. Athos da Silveira Ramos 149,
Centro de Tecnologia - Bloco C,
Cidade Universit\'aria - Ilha do Fund\~ ao,
Caixa Postal 68530, 21941-909 Rio de Janeiro,
RJ - Brasil}
\ead[url]{http://www.im.ufrj.br/$\sim$paulo/}
\ead{paulo@im.ufrj.br}

%

\begin{abstract}
We consider a continuous mathematical description of a population of ants and simulate numerically their foraging behavior using a system of partial differential equations of chemotaxis type. We show that this system accurately reproduces observed foraging behavior, especially spontaneous trail formation and efficient removal of food sources. We show through numerical experiments that trail formation is correlated with efficient food removal. Our results illustrate the emergence of trail formation from simple modeling principles.
\end{abstract}


\end{frontmatter}



%
%
%

\section{Introduction}

Ant foraging is among the most interesting emergent behaviors in the social insects. Perhaps the most striking aspect of ant foraging is how individuals following simple behavioral rules based on local information
produce complex, organized and seemingly intelligent strategies at the population level. As such, ant foraging (along with most other activities of an ant colony) is a prime example of so-called \emph{emergent} behavior.  

It has long been known that one of the main forms of communication among ants is the use of pheromones. These are chemical compounds which 
individual ants secrete and deposit on the substrate and which in effect are used as a means of communication between ants, transmitting a variety of messages such as alarm, presence of food, or providing colony-specific olfactory signatures used to identify nest-mates.

Among the many documented functions of ant 
pheromones, we are interested in their role as a chemical trail indicating the 
direction to a food source. Many species of ants, especially trail-forming 
ones, are known to lay a pheromone as they travel from the food source back to 
nest. The main attribute of this pheromone is that it is attractive to other 
ants, who tend to follow the direction of increasing concentration of the 
chemical. These ants will then reach the food source and return to the nest 
while laying pheromone themselves, thus reinforcing the chemical trail in a 
positive feedback loop. This results in the formation of well defined trails 
leading from the nest to the food source, allowing for an efficient transport 
of the food to the nest. 
Thus, pheromones play a major role in food foraging, where they are widely used (among other strategies) to recruit nest mates to new food sources.  

It is clear that the effective simulation and modeling of trail-laying and foraging behavior of ants is a crucial aspect in the understanding of ant ecology. Indeed, for invasive species such as the Fire Ant \emph{Solenopsis invicta} \cite{FireAnts} foraging is, next to reproduction, the most important activity of the colony, and the sole means by which it can ensure its nourishment. A better understanding of foraging dynamics is bound to contribute to a more complete picture of ant ecology. Aside from the scientific value of such knowledge, a thorough understanding of ant behavior is essential in defining appropriate policies in those cases (as with \emph{S. invicta} \cite{FireAnts} or the Pharaoh's ant \emph{Monomorium pharaonis} \cite{Jackson2004}) where ant species are considered pests. 

The entomological research body on ants, their behavior, and their olfactory means of communication is vast. Here, we content ourselves with citing some seminal works, as well as some more recent investigations with special relevance to our analysis. For a general reference on myrmecology (the branch of entomology that deals with ants), we refer to the encyclopedic book by H\"olldobler and Wilson \cite{TheAnts}. Therein may be found many relevant references up to 1990. The paper \cite{Regnier1968} contains an overview of the chemical study of pheromones. Concerning the trail-laying behavior of ants, and foraging in general, we refer to
\cite{Beckers1992,Bossert1963,Deneubourg1990,Edelstein,
Edelstein1995,Ramsch2012,Rauch1995,Sumpter2003,Sumpter2003-2,Kun2014,Vittori2004,
Van1986,Vowles55,Wilson1962-1,Wilson1962-2,Wilson1962-3}, and the references therein, although of course many other papers could be cited.

Concerning the computational simulation of ant trail-laying, we refer to \cite{Boissard2012,Couzin2002,Edelstein1995,Johnson2006,Ryan,Schweitzer1997,Sumpter2003,Watmough1995,Watmough1995-2,Weyer1985}, although again this list is far from complete. See especially \cite{Boissard2012} for a recent approach involving directed pheromones, and an excellent, up-to-date review of available numerical and modeling strategies for ant foraging. We encourage the reader to consult that paper for an informative discussion and overview of the state of the art in ant trail-laying simulations.

Let us just point out that, as observed in \cite{Boissard2012}, ant foraging simulations have in the past been mostly restricted to individual-based, or cellular automaton, models. That approach is certainly fruitful, but is generally limited to relatively small populations of ants, as well as somewhat restrictive modeling setups.

%

From what our bibliographical research could gather, only the work \cite{Watmough1995} presents a PDE model which (as our own) divides the ant population into two kinds, namely ants leaving the nest and ants returning to the nest (see also \cite{Johnson2006}, where the population is also divided in two different groups). However, the setting in \cite{Watmough1995} is highly simplified, being one-dimensional, so no trail formation occurs, and the system proposed in that work is only explored numerically in a simplified ODE version.

Let us also refer to the work \cite{Motta2011}, where a model for the dispersal of leaf-cutter ants is presented using PDEs. However, in that work trail-laying is not taken into account.

Thus, to the best of our knowledge, the present work is the first to consider the modeling and simulation of the whole cycle of food foraging by ants, comprising random foraging, discovery and transport of food, recruitment, formation of trails and fading of trails upon exhaustion of the food sources.\footnote{As this work went to press, we learned of the paper \cite{Bertozzi}, which proposes a system sharing many similarities with our own.}

An outline of the paper follows. In Section~\ref{Sec010}, we motivate the use of the mathematical framework of chemotaxis to model ant foraging. Next, in Section~\ref{Sec020}, we
present our modeling assumptions derived from an analysis of the myrmecological literature, and deduce our model. In Section~\ref{Sec030}, we present and discuss various numerical simulations. In Section~\ref{Var}, we perform a parameter space exploration and discuss some consequences and possible experimental validations of the model.  In Section~\ref{Sec040} we draw some conclusions from our work, discuss some of its limitations, and suggest further lines of inquiry. Finally, the Appendixes deal with the nondimensionalization procedure and the details of the numerical scheme.

The main results of this paper were announced in \cite{Amorim}.


\section{Modeling ant foraging}
\label{Sec010}

Many species of ants use recruitment of nest mates through chemical signals in order to efficiently exploit food sources. The goal is to concentrate the most individuals possible in a small region in space and time where the food source is located. This minimizes the risk of predation of the ants themselves and the removal of the food source by other foragers. To this end, eusocial insects have evolved several strategies, of which trail formation is one of the most well-known \cite[Ch.10]{TheAnts}.

Ants lay trails by depositing pheromones on the substrate, usually by pressing their sting against the substrate. Pheromones are chemical compounds that can diffuse through the substrate or through the air \cite{Bossert1963} and which the ants detect through their antennae. Importantly, ants can discern changes in concentration of the pheromone by measuring the difference in concentration between each antenna (see \cite{Boissard2012} and the references therein), thus allowing them to follow chemical gradients. A thorough description of foraging behavior for the fire ant \emph{S. saevissima} can be found in \cite{Wilson1962-1,Wilson1962-2}.

\paragraph{Chemotaxis}
In this work, we study ant foraging behavior from the mathematical point of view of \emph{chemotaxis}. The term chemotaxis is used to describe phenomena in which the movement of an agent (usually a cell or bacteria) is affected by the presence of a chemical agent. Typically, individuals follow paths of increasing (or decreasing) concentration of the chemical agent, and often produce the agent themselves. This may originate a variety of phenomena, including finite time blow-up, segregation of species, and the formation of patterns, which are of interest to biologists and mathematicians alike.

The original chemotaxis model dates back to Patlak \cite{Patlak} and Keller and Segel \cite{KellerSegel70,KellerSegel71}, and was developed to model the evolution of a density $\rho(t,x)$ of bacteria and the concentration of a chemical $c(t,x)$, $t>0$, $x\in \RR^d$, according to the system (presented here in nondimensional form) 
\be
\label{0100}
\aligned
&\del_t \rho - \Delta \rho + \dive( \rho \chi \nabla c) = 0,
\\
& \del_t c - \Delta c + \tau c = \rho,
\endaligned
\ee
with appropriate (nonnegative) initial data $\rho(0,x)$ and $c(0,x)$. The first equation models a random Brownian diffusion of the bacterial population, with a transport term with velocity vector $\nabla c$ and a sensitivity $\chi$. Thus the bacteria disperse but also have a tendency to follow the direction of steepest gradient of $c$. In turn, the second equation models the production of the chemical $c$ by the bacterial population, its diffusion on the substrate and its evaporation with rate $\tau$.
For the rich mathematical theory and a survey of results relating to chemotaxis, we refer the reader to the reviews of Hillen and Painter \cite{HillenPainter2009} and Horstmann \cite{Horstmann1,Horstmann2}.


It is natural (as had already been observed in \cite{Boissard2012}) to approach ant foraging behavior from a chemotactic point of view, where the ant population is modeled by a density function rather than by a discrete set of individuals. We will model the foraging behavior of ants by deducing a suitable generalization of the chemotaxis system \eqref{0100} to encompass two types of ant (foraging ants and returning ants), as well as the pheromone concentration and food source availability. Note that chemotactic models have already been applied successfully to multi-species situations in other settings, see for instance \cite{Painter,TangTao}.  

Finally, we note that our approach shares some aspects with mechanistic approaches to animal movement already applied to territory studies of mammal species such as wolves by Lewis and collaborators \cite{Lewis93,Lewis97,MoorcroftLewis06}. Indeed, the basic phenomena of random motion and attractiveness or repulsion to certain chemical or olfactory signals (pheromones in the case of ants, urine marks in the case of wolves) may be mathematically modeled in a similar way. 

\section{A continuous chemotaxis-type model of ant foraging}
\label{Sec020}
In this section we present our PDE model of ant foraging behavior. 
The populations of foraging and returning ants are modeled respectively by \emph{density functions} $u(t,x)$ and $w(t,x)$ depending on $x\in \Omega$ and $t\ge 0$, where $\Omega$ is an open subset of $\RR^2$, representing the physical domain that the population inhabits, and $t$ is time. Note that in the simulations below, $\Omega$ will be a bounded set, but for modeling purposes, the domain may be all $\RR^d$. Even though individual ants are not microscopic, modeling an ant population using a continuous density is reasonable. Indeed, to take the example of the genus \emph{Pogonomyrmex,} a typical worker measures 1.8mm in body length, while their foraging range usually extends to distances of 45--60\,m \cite{Holldobler1976}. Thus, the ratio of foraging distance to average body length may be of the order of $3\times 10^4.$ The same assumption is used frequently, for instance, in the continuous modeling of cell dynamics \cite{HillenPainter2009}. 

Moreover, as is customary when modeling physical phenomena using reaction--diffusion equations, such as crowd movements or chemotaxis, the solution may be seen as the averaged outcome of a great number of individual experimental runs.


\subsection{Presentation of the model}
We propose the following continuous model for ant foraging, which will be explained in subsequent sections: 

\be
\label{500}
\left\{
\begin{aligned}
&\partial_t u -   \alpha_u \Delta u + \dive\big ( u\, \beta_u \nabla v   \big) = - \lambda_1 uc + \lambda_2 w \overline N(x) + \overline M(t) \overline N(x)
\\
& \partial_t w - \alpha_w \Delta w + \dive\big ( w\, \beta_w {{\nabla a}} \big) = \lambda_1 uc - \lambda_2 w \overline N(x)
\\
& \partial_t v = \mu\overline P(x) w - \delta v + \alpha_v \Delta v
\\
& \partial_t c = - \gamma u\, c.
\end{aligned} \right.
\ee

The variables and quantities used in our model have the following meanings:
\medskip
$$
\begin{tabular}{r  l}
Variables &  \\ [0.5ex] 
\hline\\ [0.1ex] 
$u(t,x)$ &  density of foraging ants  \\ [0.5ex] 
$w(t,x)$ &  density of ants returning to the nest with food  \\ [0.5ex] 
$v(t,x)$ &  concentration of pheromone  \\ [0.5ex] 
$c(t,x)$ &  concentration of food source  \\ [0.5ex] 
Given functions &  \\ [0.5ex] 
\hline\\ [0.1ex] 
${{\nabla a}}(x)$ &  nest-bound field  \\ [0.5ex] 
$\overline N(x)$ & describes location of the nest \\ [0.5ex] 
$\overline M(t)$ & describes foraging ants emerging from the nest  \\ [0.5ex] 
$\overline P(x)$ & describes decrease in pheromone deposition near the nest.  \\ [0.5ex] 
\end{tabular}
$$
\medskip	

The constant parameters appearing in system \eqref{500} are collected in Table~\ref{T10} below, while
the quantities appearing in \eqref{500} are the following given functions: $\nabla a$ is a  given nest-bound vector field, so that $a(x)$ is (the negative of) a potential-like function indicating distance to the nest.
$\overline N(x),$ which describes the spatial placement of the nest, in such a way that $\int_\Omega \overline N(x) \,dx$ gives the total area of the nest entrance; more exactly, the support of $\overline N(x)$ represents the small region around the nest entrance where returning ants turn into foraging ants. 

The function $\overline M(t)$  is of the form $C \chi_{(0,T)}$ for some $C,T>0$ (here $\chi$ denotes a characteristic function) and describes the foraging ants emerging from the nest at rate $C$ until time $T$; and $\overline P(x)$, which is a function which decreases to zero as one approaches the nest, intended to reflect the experimental fact that ants decrease pheromone deposition as their distance to the nest decreases. We take $\overline P$ with a paraboloid profile in the simulations below.

\subsubsection*{Physical parameters}

In Table \ref{T10}, we collect the physical parameters intervening in system \eqref{500}.
Note that we do not provide estimates for the values of these values here, primarily since the variations among ant species are huge. Furthermore, many of them are actually rather difficult to obtain in the literature. Moreover, the nondimensionalisation in Section~\ref{nondim} below reduces the number of parameters of which we must know an exact value, so that it is only necessary to know certain ratios between them. 

\begin{table}[htpb]
\caption{Physical parameters in system \eqref{500}. Here, $\ell$ denotes length, $t$ denotes time, $\mathop{food}$, $\mathop{phero}$ and $\mathop{ants}$ denote some measure of, respectively, food, pheromone, and ant quantity.\label{T10}}
\begin{tabular}{p{0.2\textwidth}|p{0.2\textwidth} | p{0.5\textwidth}}
Parameter & Units & Physical meaning \\ \hline
\hline&\\[-1.8ex]  
$\alpha_u, \alpha_w $ & $\ell^2/t$ &  Diffusion rate of foraging and returning ants  \\ [0.5ex] 
$\beta_u $& $\dfrac{\ell^4}{t\cdot\mathop{phero} }$ &  Foraging ants' pheromone sensitivity   \\ [2.5ex] 
$\beta_w $ & $\ell/t$ &  Returning ants' sensitivity to the nest-bound field ${\nabla a}(x)$   \\ [0.5ex] 
$\lambda_1  $ & $\dfrac{\ell^2}{t\cdot \mathop{food}} $ & Rate of transformation of foraging into returning ants at food site  \\ [0.5ex] 
$\lambda_2 $ &  $t^{-1} $ & Rate of transformation of returning ants into foraging ants at nest    \\ [0.5ex] 
$\mu $ & $\dfrac{\mathop{phero}}{t\cdot\mathop{ants}} $ &  Rate of pheromone deposition   \\ [2.5ex] 
$\delta $ & $t^{-1}$ & Rate of pheromone evaporation   \\ [0.5ex] 
$\alpha_v $ & $\ell^2/t$ & Rate of pheromone diffusion   \\ [0.5ex] 
$\gamma$& $\dfrac{\ell^2}{t \cdot\mathop{ants}}$ & Rate of food removal by foraging ants   \\ [2.5ex] 
%
\hline
\end{tabular}
\end{table}

The system \eqref{500} must be supplemented with appropriate boundary conditions and initial data.

\subsubsection*{Boundary conditions and initial data}

In the interest of conservation of the total mass of ants,
we impose zero-flux boundary conditions on system~\eqref{500}. That is,  we assume that on the boundary $\del\Omega$, it holds
\be
\label{2500a}
\aligned
&\big( \alpha_u \nabla u - \beta_u u\nabla v  \, \big) \cdot \nn = 0,
\\
& \big ( \alpha_w \nabla w  -  \beta_w w {\nabla a}   \big) \cdot \nn = 0,
\\
& \nabla v \cdot \nn = 0, \qquad c = 0,
\endaligned
\ee
where $\nn$ is the outward unit normal vector to $\del\Omega$.

Note that imposing the boundary conditions \eqref{2500}, one can easily check that, at least for smooth solutions, the system \eqref{2000} preserves the total mass of ants, after they have all emerged from the nest: for each $t,$
\[
\aligned
\int_\Omega u(t) + w(t) \,dx = C,
\endaligned
\]
where $C$ is the total quantity, or mass, of ants.

The initial data are
\be
\label{2600a}
\aligned
u(0,x) = w(0,x)= v(0,x) = 0, \qquad c(0,x)= c_0(x).
\endaligned
\ee
At $t=0$, no ants and no pheromone are present. Foraging ants will emerge from the nest according to the function $\overline M(t)$ in \eqref{500}, as described above.
An initial distribution of food is provided by the function $c_0$. 

\subsubsection*{Analytical results}
Mathematically, the system \eqref{500}--\eqref{2600a} is of chemotaxis type and so is ame\-nable to rigorous analysis. In fact, well-posedness results are available for this system, which will be the object of a forthcoming paper \cite{AAG}.
Another important property of the system \eqref{500}--\eqref{2600a} worth mentioning here is that the time evolution preserves the positivity of solutions, see \cite{AAG}. 

As we shall see in Section \ref{Var} below, for certain parameter ranges, the formation of trails is weak or nonexistent. The well-posedness results in \cite{AAG} do not take this parameter dependence into account. Thus, it would be interesting to see whether any rigorous estimates of the parameter ranges allowing trail formation could be obtained.

\subsubsection*{Reference density}
At this point we introduce a reference density $\uref$, which emerges naturally from the boundary conditions \eqref{2500a}. It is such that, after all the ants have emerged from the nest, $(u,w,v,c)= ( \uref, 0,0,0) $ is a steady-state solution to the system \eqref{500}. Since the total mass of ants $C$ is conserved eventually, we must have $\uref = C/ |\Omega|$.

\subsubsection*{Summary of the model}
The system \eqref{500} may be interpreted as follows:
\begin{itemize}
\item the foraging $u$-ants emerge from the nest located at $x=0$ at rate $\overline M$ until time $T$ (or else their initial distribution may be a given function). They disperse randomly according to Fourier's heat law (or Fick's law) ($\alpha_u \Delta u$ term).
Upon encountering the pheromone $v$, they follow its gradient ($\dive (u \beta_u\nabla v)$ term). 

\item When they reach an area with food (indicated by a positive food concentration 
$c$), foraging ants turn into $w$-ants (returning ants), by means of the coupling source terms $\pm\lambda_1 uc$, depleting the food in the process (which is described by the fourth equation). 
\item The $w$-ants now return to the nest with food, by following the (prescribed) nest-bound vector field ${\nabla a}(x)$. 
This is modeled by the transport term $\dive\big ( w\,   \beta_w {{\nabla a}} \big).$
They lay the pheromone $v$ which the $u$-ants will now follow to reach the food, according to the chemotactic transport term $\dive( u\, \beta_u \nabla v)$ in the first equation.
\item When the $w$-ants reach the nest whose location is modeled by the function $\overline N(x)$, they leave the food at the nest and re-emerge as $u$-ants. 
\item The third equation represents the laying of the pheromones by the $w$-ants, which evaporates and diffuses. Note the function $\overline P(x)$, describing the decrease in pheromone deposition as ants approach the nest.
\item The last equation represents the depletion of the food when foraging ants come in contact with it.
\end{itemize}

\medskip

\subsection{Modeling assumptions}
\label{Mod}

We now describe the modeling assumptions leading to the formulation of system \eqref{500},\eqref{2500a}.
We intend to model not one specific species of ant, but rather to capture in a qualitative way the characteristic properties of ant foraging. In view of this, we will borrow behavioral aspects from various species of ants. However, in the simulations below we must use concrete values for the various physical quantities involved, and for this we shall rely on various sources. We use some experimental data on \emph{Lasius niger}, collected in \cite{Boissard2012} and available also in \cite{Beckers1992}. Mostly, though, we rely on the descriptions in Wilson's works \cite{Wilson1962-1,Wilson1962-2} on \emph{S. saevissima}.

%
%

\paragraph{Returning to the nest}
One of the main assumptions of our model is that ants know the way back to their nest upon finding a food source. This is reflected by the introduction of a given nest-bound vector field $\nabla a(x)$, derived from a potential-like function $a(x)$ which in the simplest case is just the negative of the distance to the nest (smoothed at the nest site). Importantly, this assumption is supported by the literature. Indeed, many species of ants have been proven to use visual and olfactory orientation cues to return to the nest \cite{Holldobler1976,Steck2009}, as well as so-called orientation by path integration \cite{Muller1988,Wehner03}, in which individuals cumulatively keep track of changes in direction and thus of the overall direction of the nest. Even the concentration of carbon dioxide (which is greater near the nest) is conjectured to serve as a homing guide for returning ants, see \cite[p.289]{TheAnts}.

\paragraph{Dynamics near the nest and at the food site}
We will assume that the nest is a small but extended region in which ants returning to the nest carrying food are transformed into foraging ants at a certain rate. We do not take into account any eventual time ants might spend inside the nest. Rather, we suppose that upon reaching the nest entrance, ants instantaneously drop their food and return to foraging. Although this might not be realistic, we make the simplifying assumption that the mass of foraging ants inside the nest is negligible.

We make no attempt to model the mechanics of recruitment which take place in the nest or near its entrance, by which returning ants recruit other ants, often by antenna contact or by physically pushing them in the direction of the trail. See \cite[p.279]{TheAnts} for a description of some species' intricate dynamics near the nest, and \cite{Kun2014} for a model of these dynamics, with an analysis of the measure in which they may influence foraging.

At the food site, an inverse transformation takes place: foraging ants are transformed into returning ants at a fixed rate. Here, we suppose that the ants spend no time feeding at the food site and are able to very quickly grab a portion of food and start their journey back to the nest. Although this is also probably an oversimplification, the resulting  equation for the removal of food by the ants is extremely convenient, not least because it allows one to define the ``half-life'' of the food (that is, the mean time it takes ants having a certain reference density to remove half the available food), which will serve as the time scale used in the nondimensionalisation procedure below.


\paragraph{Choice of direction when encountering a trail}
In line with the experimental results in \cite{Beckers1992}, we will suppose that the rate of pheromone deposition decreases as ants approach the nest. This experimental fact, observed at least in \emph{L. niger} ants, is especially convenient from the modeling point of view since it acts to prevent over-concentration of pheromone near the nest. 

This is related to the more general problem individual ants face when encountering a trail, of deciding in which direction to follow the trail. Again our study of the literature yields mixed results. For instance, in \cite{Wilson1962-1} it is reported that ants encountering a trail immediately follow it in the direction \emph{away} from the nest. This is consistent with our assumption that individuals know the general direction of the nest. However, in the same work instances are reported where an individual ant returning from a food source laying pheromone is followed closely by another forager, who is obviously attracted to the pheromone, but is traveling in the ``wrong'' direction. Also, frequent double-backs are reported in this and other works, and these may even provide a way to reinforce the trail when in its early stages \cite{Reid2012}.    
A more recent study \cite{Jackson2004-2} shows that at least for the Pharaoh’s ant \emph{M. pharaonis}, the geometry of bifurcations of the trail serves as an indicator of the polarity of the trail (i.e.,~of its ``right'' direction): individuals traveling in the wrong direction correct their path when encountering a bifurcation by analyzing the angles between trail branches.

%
%

The gradual decrease in pheromone deposition when approaching the nest reported in \cite{Beckers1992} actually provides a plausible and simple mechanism allowing the ``correct'' food-bound direction to be chosen more often when an individual encounters a trail. Indeed, as observed in the simulations below, the resulting pheromone profile presents a clear slope leading away from the nest, which is \emph{not} orders of magnitude smaller than the slope in pheromone concentration transverse to the trail, at least for the relatively short trails we simulate. This allows the ants to follow the trail in the direction of the food. 

It would be interesting to know if the decrease in pheromone deposition when approaching the nest is common to other species, and whether it has any relation to the choice of a direction when following a trail. Our numerical results suggest it does, although of course results from a model as simplified as ours must be interpreted with caution, and trail polarity results such as \cite{Jackson2004-2} must be taken into account.

\paragraph{Modeling pheromone propagation}
\label{phero}
The modeling of pheromone propagation is not straightforward. First, observe that we suppose the ants move in a two-dimensional domain in which they deposit pheromone, which diffuses. But the pheromone will actually diffuse through the half-space of air above the plane of the ants. Therefore, strictly speaking, it should obey a \emph{three-dimensional} diffusion equation on a half-space, with initial data concentrated on its boundary (where the ants deposit the pheromone). It is not clear from the literature whether there is any strictly two-dimensional diffusion along the substrate. 

Fortunately, the classical work by Bossert and Wilson \cite{Bossert1963} provides some guidelines. The authors measured and simulated pheromone diffusion using a variety of approaches, taking into account the observations made above, and suggested that one can model its propagation according to a two-dimensional diffusion process acting on the substrate, and provide actual estimates for the diffusion coefficient, which we will use.

Thus, we propose a two-dimensional diffusion equation of the type
\be
\label{0110}
\aligned
\partial_t v = \mu w - \delta v + \alpha_v \Delta v
\endaligned
\ee
to model pheromone diffusion.

In our case, $- \delta v$ models only the chemical degradation of the pheromone. According to \cite[p.244]{TheAnts} and \cite{Bossert1963}, the degradation rate should be quite high to allow for quick abandonment of non-productive trails; still, in \cite{Regnier1968} a great variation in pheromone degradation speed is observed, with trails remaining detectable from a few hundred seconds to days. See also \cite{Jeanson2003} for specific values of pheromone trail decay rates in the case of \emph{M. pharaonis}. Another advantage of using an equation of type \eqref{0110} is that one remains inside the well-studied framework of chemotaxis. We will see that, despite these caveats, the choice of \eqref{0110} for pheromone dynamics provides good results in our framework.


\subsection{Nondimensional system}
\label{nondim}

In  \ref{nondim}, we deduce the following non dimensional version of the system \eqref{500},
\be
\label{2000}
\left\{
\begin{aligned}
&\partial_t u -  \Delta u + \dive\big( u\, \chi_u\nabla v  \big) = - uc + \lambda w N(x) + M(t) N(x)
\\
& \partial_t w - D_w \Delta w + \dive\big( w\,  {\nabla a}   \big) =  uc - \lambda w N(x)
\\
& \partial_t v = P(x) w - \varepsilon v +  D_v \Delta v
\\
& \partial_t c = - u\, c.
\end{aligned} \right.
\ee
The boundary conditions and initial data \eqref{2500a}, \eqref{2600a} become

\be
\label{2500}
\aligned
&\big(  \nabla u - \chi_u u\, \nabla v   \big) \cdot \nn = 0,
\\
& \big ( D_w \nabla w  -  w\, {\nabla a}  \big) \cdot \nn = 0,
\\
& \nabla v \cdot \nn = 0, \qquad c = 0,
\endaligned
\ee
where $\nn$ is the outward unit normal vector to $\del\Omega$, and
\be
\label{2600}
\aligned
u(0,x) = w(0,x)= v(0,x) = 0, \qquad c(0,x)= c_0(x).
\endaligned
\ee

The original system \eqref{500} is thus reduced to system \eqref{2000}, having the dimensionless parameters and given functions described in Table~\ref{T20}.

\begin{table}[htpb]
\caption{Dimensionless parameters in system \eqref{2000}\label{T20}}
\begin{tabular}{ p{0.4\textwidth}|  p{0.5\textwidth}}
Dimensionless parameter & Physical meaning \\ \hline
\hline&\\[-1.8ex] 
$\varepsilon= \delta/(\gamma \uref)$ & Pheromone degradation rate relative to the time-scale $\hat t$	  \\ [0.5ex]  
$D_v = \alpha_v / \alpha_u  $ &  Ratio between diffusion coefficients of pheromone  and foraging ants   \\ [0.5ex] 
$ D_w= \alpha_w / \alpha_u  $ &  Ratio between diffusion coefficients of returning ants and foraging ants \\ [0.5ex] 
$\chi_u = (\beta_u\mu)/(\alpha_u\gamma)$ &  Foraging ants' pheromone sensitivity   \\ [0.5ex] 
$\lambda = {\lambda_2}/({\gamma \uref}) $ & Rate of transformation of returning ants into foraging ants at nest, relative to the time-scale $\hat t$	  \\ [0.5ex] 
$P(x)$ & Describes decreasing pheromone deposition when close to the nest   \\ [0.5ex] 
$M(t) $ & Describes foraging ants emerging from the nest \\ [0.5ex] 
$N(x)$  & Location of the nest \\ [0.5ex] 
$a(x)$  & Potential-like function describing attraction to the nest \\ [0.5ex] 
\hline
\end{tabular}

\end{table}

\section{Numerical simulation of ant foraging behavior}
\label{Sec030}

In this section we present several numerical integrations of the nondimensional system \eqref{2000}--\eqref{2600}. Details of the numerical procedure may be found in \ref{numerical}.

\subsection{Trail formation with two food sources}
We simulate a setting  similar to the one reported in Wilson's experiments with the fire ant \emph{S. saevissima} \cite{Wilson1962-1,Wilson1962-2,Wilson1962-3}. The domain is a square arena of $4\, \mathrm{m}^2$, with the nest at its center, in which two separate food sources are placed at a distance of approximately $100\,\cm$ from the nest.

We emphasize that our simulations are not intended to precisely simulate a particular species, but rather to illustrate the emergence of trail formation from simple modeling principles.

The discretization comprises a $160 \times 160$ point grid with a total of $25\,600$~points, and a time-step of $0.002$ was used, which corresponds to $0.204\sec$. The space increment is $dx = dy = 0.225$ in the units of \eqref{2000} (see \ref{nondim}), which corresponds to $dx=dy= 1.25 \cm$.

\subsubsection*{Estimating parameters}
\label{Sec5000}

We now turn to the question of estimating actual values for the parameters and functions appearing in Table~\ref{T20}. We will see it is not straightforward to determine realistic values for every parameter, partly due to a lack of precise estimates in the literature. 


Let us begin with the values that are well-established in the literature. 
First, we will take a colony size of $75\,000$ ants. Colony size is highly variable \cite[p.160]{TheAnts}, varying from less than 10 to a few million. In an area of 4 $\mathrm{m}^2$ (a typical value, for instance, for a young fire ant colony \cite{FireAnts}, or for an experimental setting), this gives a reference density $\uref = 1.875$ ants per $\cm^2$. 

\paragraph{Diffusion}
We will use for the pheromone diffusion coefficient the value $\alpha_v = 0.01 \frac{\cm^2}{\seg}.$ This is the value used in \cite{Boissard2012}, based on the slightly lower value determined in \cite{Bossert1963} (which the authors claim is probably underestimated). Surprisingly, we could not find an accurate value for the foraging ants' diffusion coefficient $\alpha_u$ in the literature. To determine the nondimensional parameter $D_v$ (see Table~\ref{T20}, we assume that foraging ants' diffusion coefficient $\alpha_u$ is ten times larger than $\alpha_v$. This is supported by values estimated in \cite{Motta2011}, for leaf-cutter ants of the genus \emph{Atta}, who propose $\alpha_u = 0.39 \frac{\cm^2}{\seg}$. Since leaf-cutter ants are large, fast ants, we lower this value a little to $\alpha_u = 0.1 \frac{\cm^2}{\seg}$  and will thus use $D_v= 0.1$. 

The returning ants have no advantage to divert from their path by random movement, and so we assume that their diffusion coefficient is the same as the pheromone's (i.e., small), thus giving $D_w=0.1$.

\paragraph{Time and space scales}
We may now estimate $\hat t$ and $\hat x$. 
Recall that $\hat t = (\gamma \uref)^{-1}$. This allows us to relate it to the rate of food removal in the following way. The last equation in the original system \eqref{500} is $\del_t c = -\gamma uc$. We focus our attention on a single point in space and suppose temporarily that $u \equiv \uref$ on that point. In these circumstances, the half-life of the food could be measured experimentally, which as far as we know has not been done (assuming, of course, that ants remove the food according to the law appearing in \eqref{2000}; this may not be accurate, see the discussion in the Conclusions section at the end of this paper). We are left with positing a reasonable value for this half life, which we set at $70\, \seg$.
A short calculation then gives $\hat t = (\gamma \uref)^{-1} = 70 \,\seg/\ln2 \simeq 102\, \seg$. The spatial scale is then determined as $\hat x = \sqrt{\alpha_u \hat t} \simeq 5.5\,\cm$.

Note that the choice of $\hat t$ by this type of reasoning contains assumptions on the amount of food a single ant can carry. Indeed, for the same reference ant density $\uref$, a shorter food half-life (and thus different $\hat t$) must mean that the same number of ants can carry away more food in the same time. 

\paragraph{Pheromone degradation and sensitivity}
We take the value associated to pheromone degradation to be $ \varepsilon =0.5$, for the sake of definiteness. This is obtained by assuming that the pheromone half-life is $140\, \seg$. Note that, as observed in \cite{Regnier1968}, trail pheromones degradation speeds vary widely, with trails remaining detectable from a few minutes to days. Here we take a value giving a rather long time of trail degradation, as suggested in \cite{Wilson1962-1}.

The pheromone sensitivity $\chi_u$ is probably the most difficult parameter to accurately estimate. 
We set a value of $\chi_u = 50$ for the sake of illustration. As described in the numerical experiments of Section~\ref{Var}, we have simulated the system \eqref{2000} for a variety of values of $\chi_u$ and so refer to that section for more details.

\paragraph{Food-related parameters}
To estimate $\lambda$, consider that the speed of transformation of returning ants into foraging ants should be quite high, to prevent clogging at the nest. Therefore, assuming a density (using the variables of system \eqref{2000}) of $w=1$ at the nest, and a small ``half-life'' of returning ants at the nest, we set $\lambda\simeq 70$.

%
Let us consider the simplified system
\be
\label{2800}
\aligned
\del_t u = -uc
\\
\del_t c = -uc,
\endaligned
\ee
which isolates the dynamics of food removal, and
assume temporarily that $u$ and $c$ are only time-dependent.
Since system \eqref{2800} reduces to the standard logistic equation, one can easily check 
%
%
that, disregarding spatial dynamics, if at some time the food concentration is greater than the ant concentration, then the system \eqref{2800} will evolve to a steady state with a positive amount of food and no ants; this corresponds to a situation where ants are not very efficient, or the food is not desirable. If, in contrast, the ant density is greater that the food density in this scaling, then all the food will be removed (asymptotically) and some ants will still remain. This reflects a greater efficiency in food removal, or a great desirability of the food.


These remarks suggest that taking initial data $c_0$ with numerical value greater than the reference ant density $\uref$  will model a situation in which food removal efficiency (or food desirability) is low, and smaller values of $c_0$ model a situation in which food removal is very rapid (or food desirability is high). So, the choice of $c_0$ reflects ant efficiency and food desirability. For this simulation, we take $c_0$ concentrated on two small regions with maximums of 12 and 6, which means desirable food sources.

\paragraph{Choice of functional terms}
We now turn to the choice of the functions $P,N,M$ and $a$ in \eqref{2500}. 
The function $N(x)$ is a smoothed characteristic function indicating the nest entrance; we will suppose that the nest entrance is a circle of radius $ 10\,\cm$ situated at $x=(0,0)$ (see the discussion on the dynamics near the nest in a previous section). 

$P(x)\in[0,1]$ will be a smooth function of the form 
$ C x^2$
so that pheromone deposition is progressively reduced near the nest, which is at the origin. The constant $C$ is adjusted in each simulation to allow this reduction to become noticeable at a distance of about half the typical trail length from the nest.

To define $M(t)$ (modeling the emergence of foraging ants from the nest at the start of the simulation), we make the following assumptions. $M(t) = C_M \chi_{[0,T]}$ for some $C_M,T$, where $\chi_{[0,T]}$ denotes the characteristic function. Then, supposing that ants emerge from the nest at a rate of one ant per $\cm^2 \seg$, that the total population is about $75\,000$ ants, and converting to the new units, we obtain the appropriate values of $C_M$ and $T$.

The nest-bound potential $a(x)$ is a smoothing of the function $-C|x|$. This way, the vector field $\nabla a$ points to the nest with constant norm, except near the nest where it decays to zero. Therefore, $C$ must be the value (in $\hat x / \hat t$ units) of an individual ant's speed returning to the nest, which we suppose is $1 \cm/\seg = 18.35\, \hat x / \hat t$. Thus we take $a(x) = 18.35 |x|$ in \eqref{2000}.

We collect in Table \ref{T40} the actual values used in this simulation.

\subsection{Numerical results of trail formation with two food sources}
\begin{figure}[htbp]
$$
\hspace{-0mm}\includegraphics[trim=80 80 10 50,clip,width=.33\textwidth]{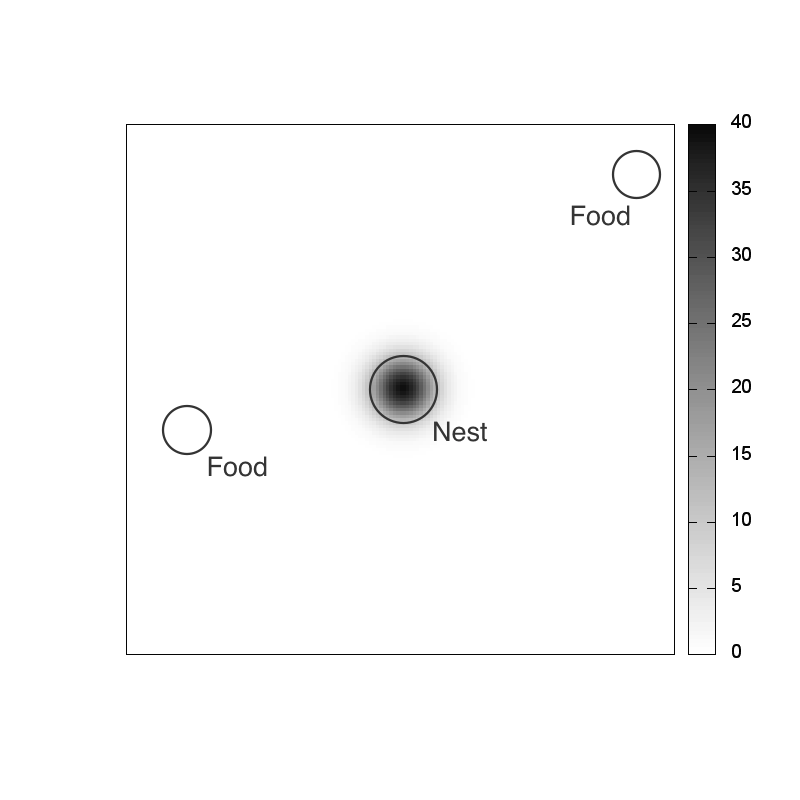}
\hspace{-0mm}\includegraphics[trim=70 70 10 50,clip,width=.33\textwidth]{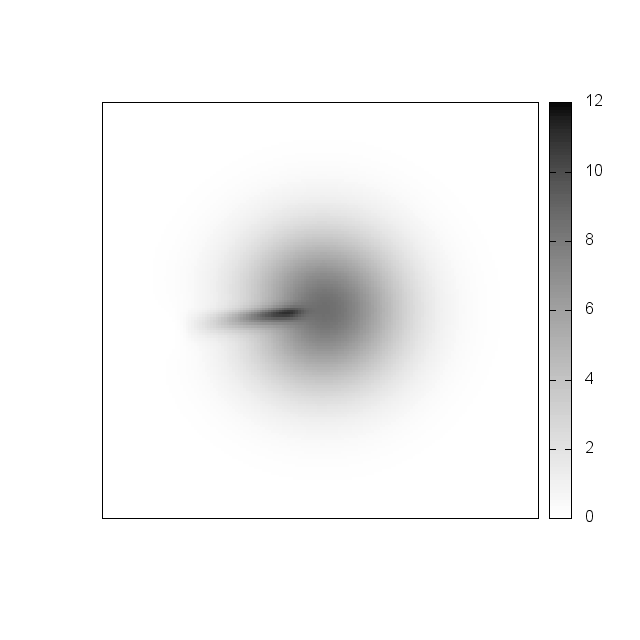}
\hspace{-0mm}\includegraphics[trim=70 70 10 50,clip,width=.33\textwidth]{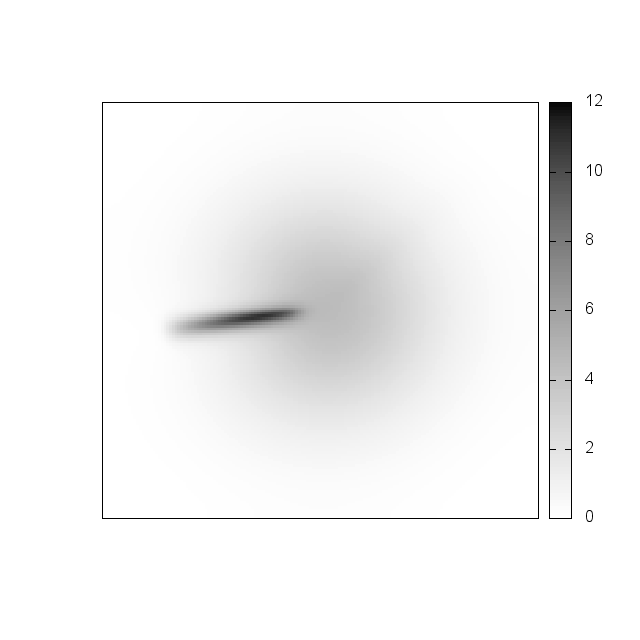}
$$
$$
\hspace{-0mm}\includegraphics[trim=70 70 10 50,clip,width=.33\textwidth]{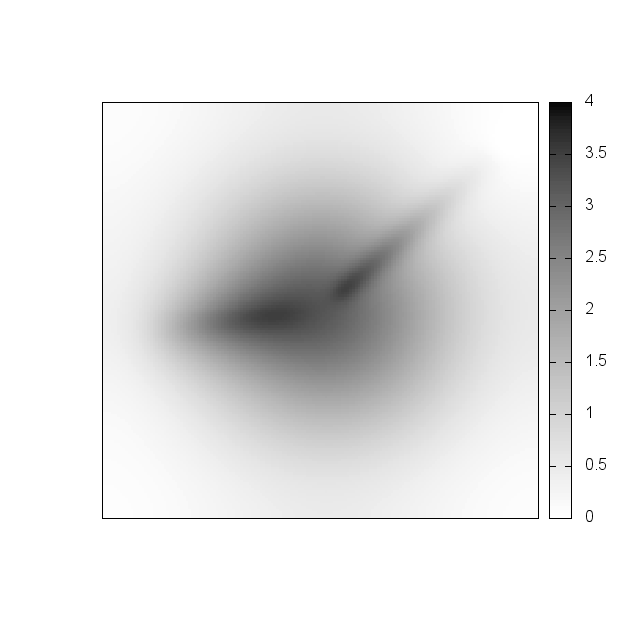}
\hspace{-0mm}\includegraphics[trim=70 70 10 50,clip,width=.33\textwidth]{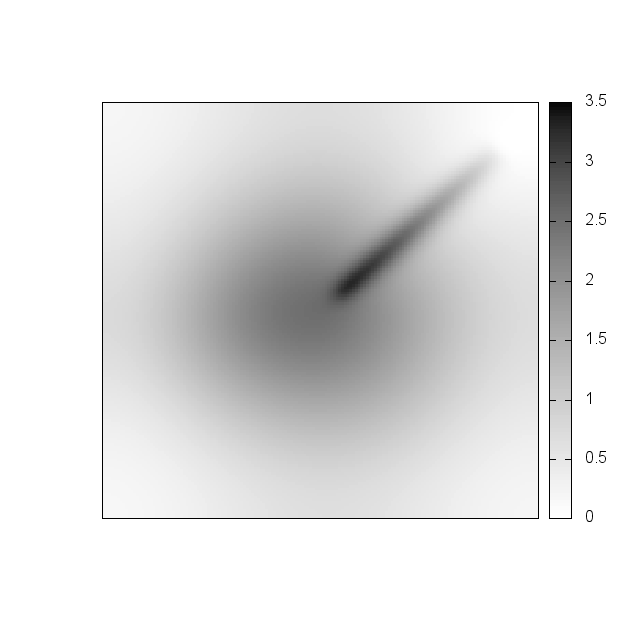}
\hspace{-0mm}\includegraphics[trim=70 70 10 50,clip,width=.33\textwidth]{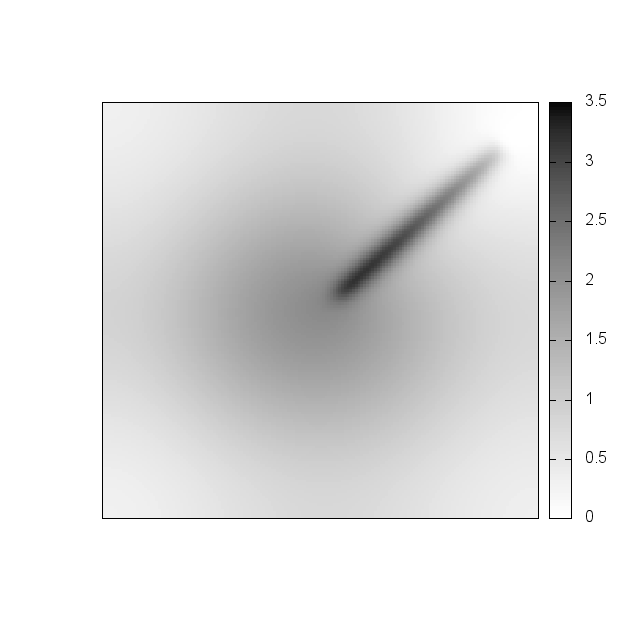}
$$
$$
\hspace{-0mm}\includegraphics[trim=70 70 10 50,clip,width=.33\textwidth]{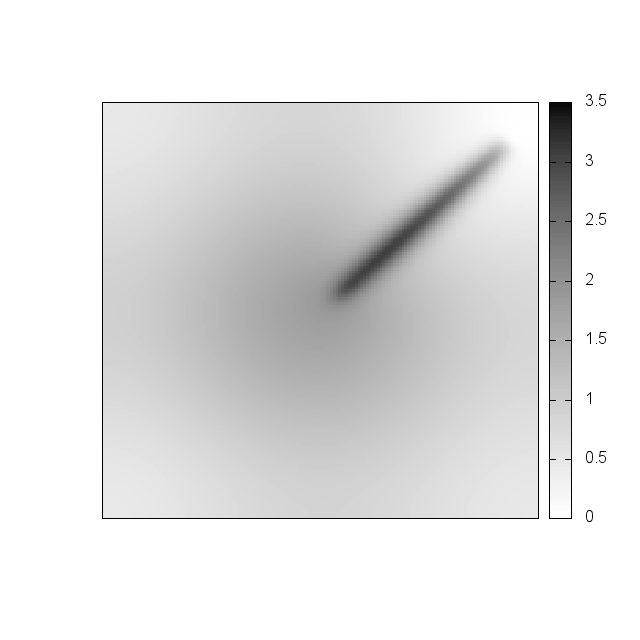}
\hspace{-0mm}\includegraphics[trim=70 70 10 50,clip,width=.33\textwidth]{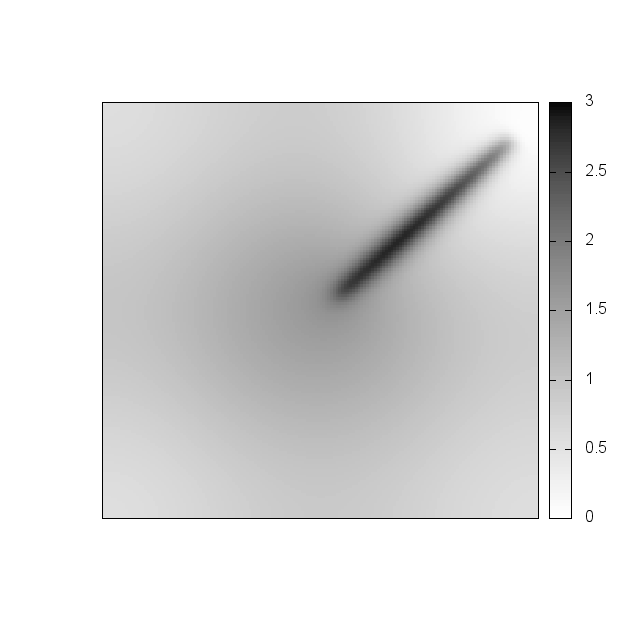}
\hspace{-0mm}\includegraphics[trim=70 70 10 50,clip,width=.33\textwidth]{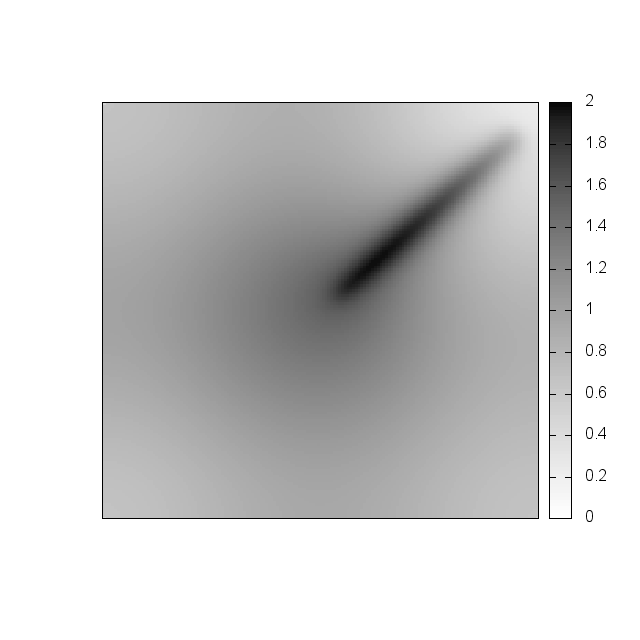}
$$
$$
\hspace{-0mm}\includegraphics[trim=70 70 10 50,clip,width=.33\textwidth]{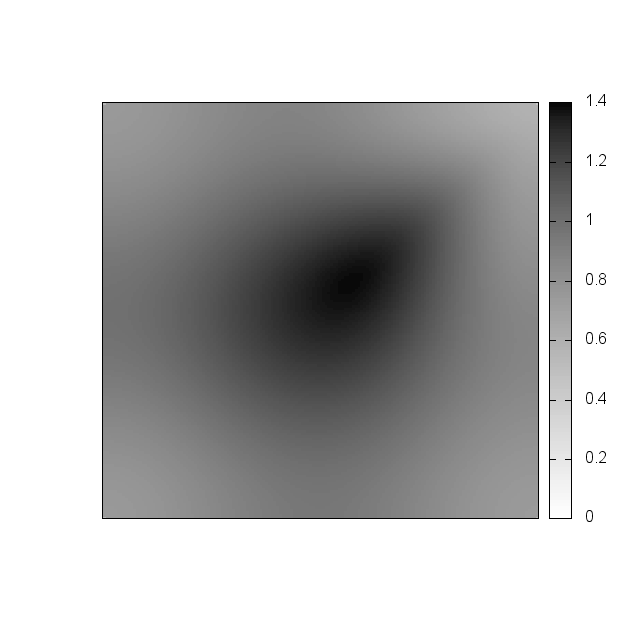}
\hspace{-0mm}\includegraphics[trim=70 70 10 50,clip,width=.33\textwidth]{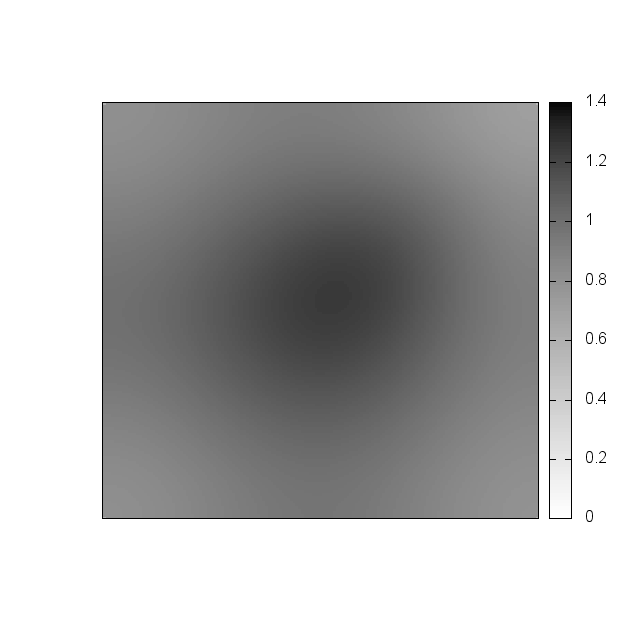}
\hspace{-0mm}\includegraphics[trim=70 70 10 50,clip,width=.33\textwidth]{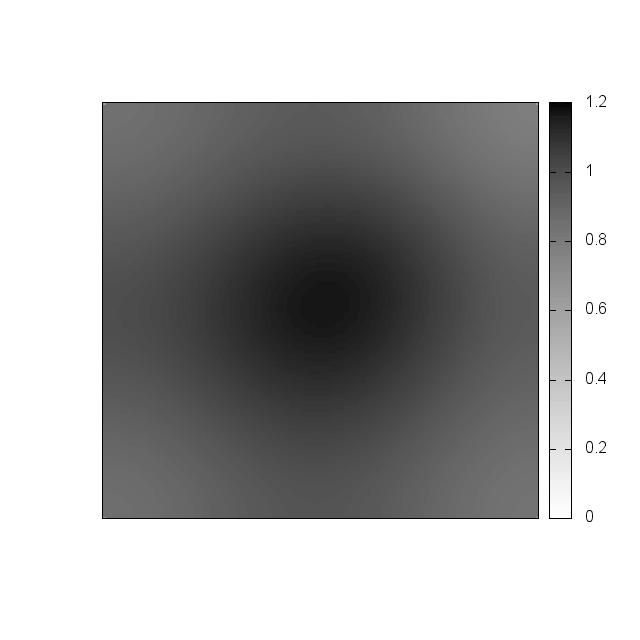}
$$
\caption{\small Foraging ant density (left to right, top to bottom) from $t= 14 \min$ in increments of $14 \min$, up to the final time of 2.8 hours. Domain is $200\,\cm \times 200 \, \cm.$}
\label{FigU1}
\end{figure}

\begin{figure}[htbp]
$$
\hspace{-0mm}\includegraphics[trim=80 80 10 50,clip,width=.33\textwidth]{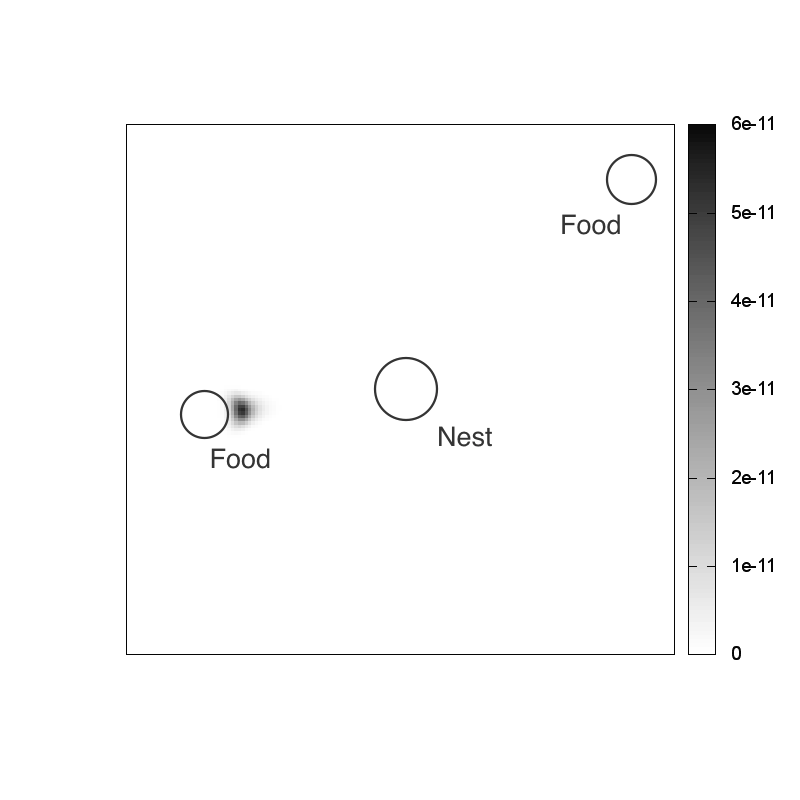}
\hspace{-0mm}\includegraphics[trim=70 70 10 50,clip,width=.33\textwidth]{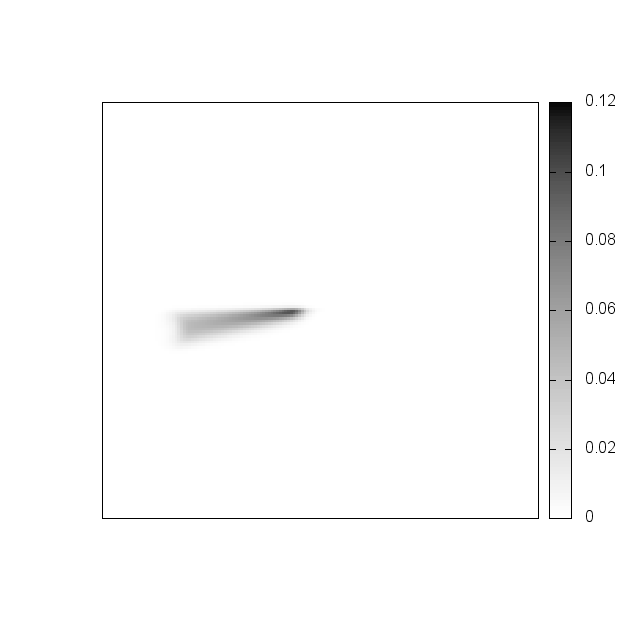}
\hspace{-0mm}\includegraphics[trim=70 70 10 50,clip,width=.33\textwidth]{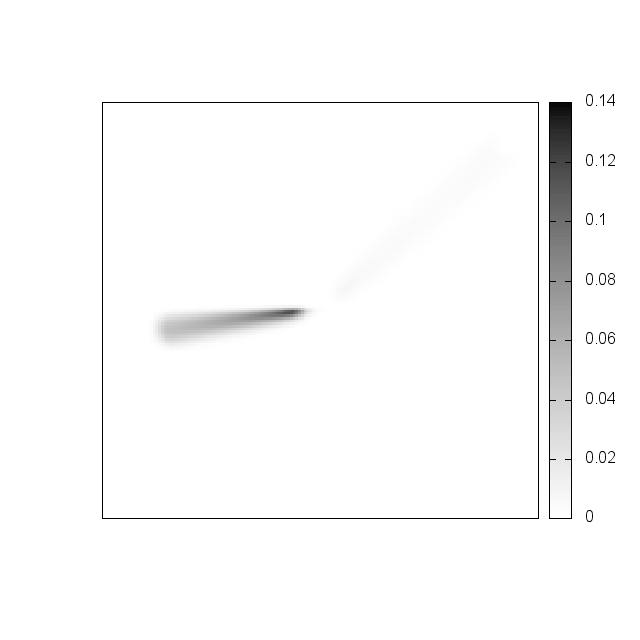}
$$
$$
\hspace{-0mm}\includegraphics[trim=70 70 10 50,clip,width=.33\textwidth]{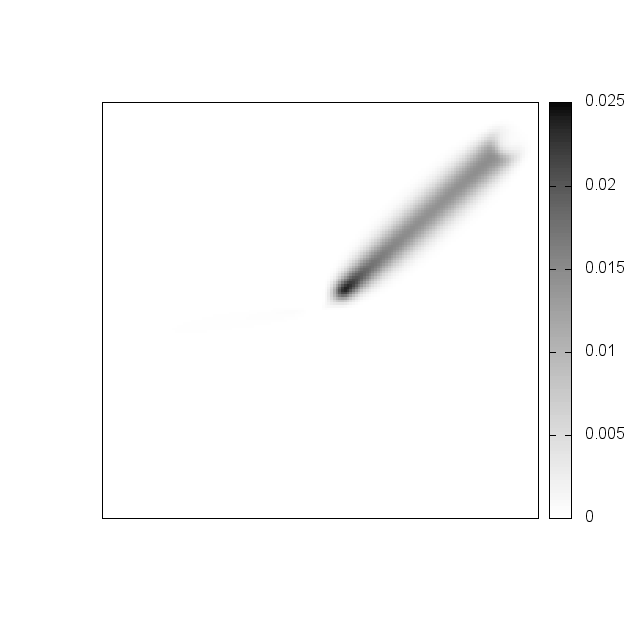}
\hspace{-0mm}\includegraphics[trim=70 70 10 50,clip,width=.33\textwidth]{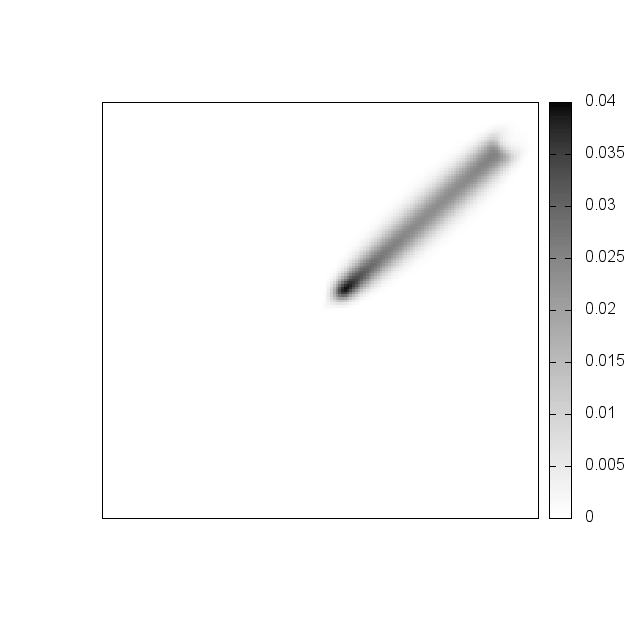}
\hspace{-0mm}\includegraphics[trim=70 70 10 50,clip,width=.33\textwidth]{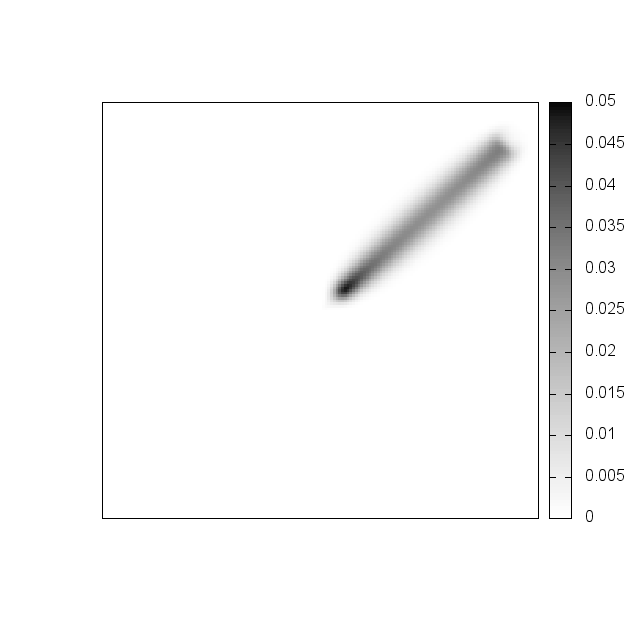}
$$
$$
\hspace{-0mm}\includegraphics[trim=70 70 10 50,clip,width=.33\textwidth]{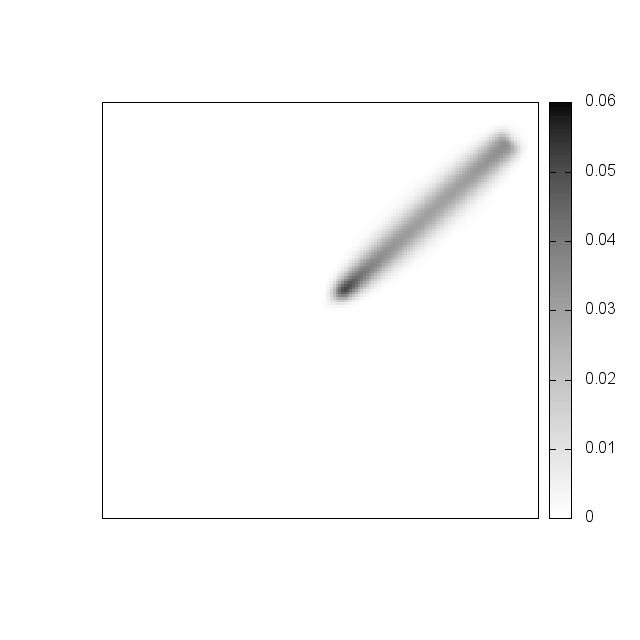}
\hspace{-0mm}\includegraphics[trim=70 70 10 50,clip,width=.33\textwidth]{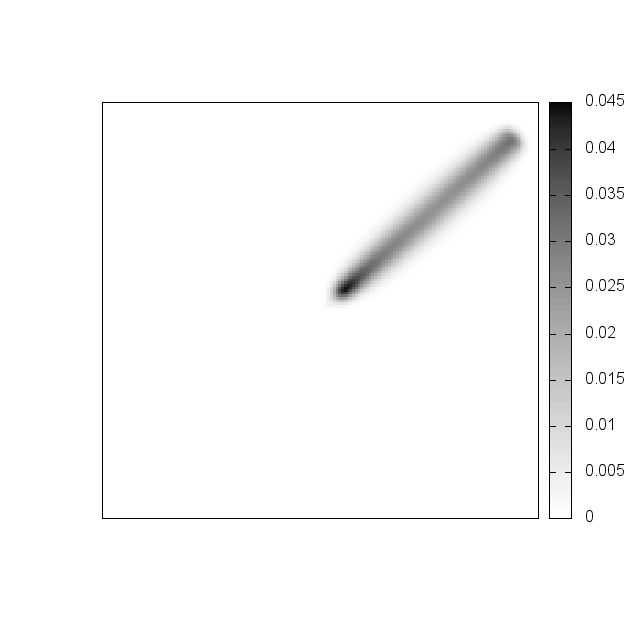}
\hspace{-0mm}\includegraphics[trim=70 70 10 50,clip,width=.33\textwidth]{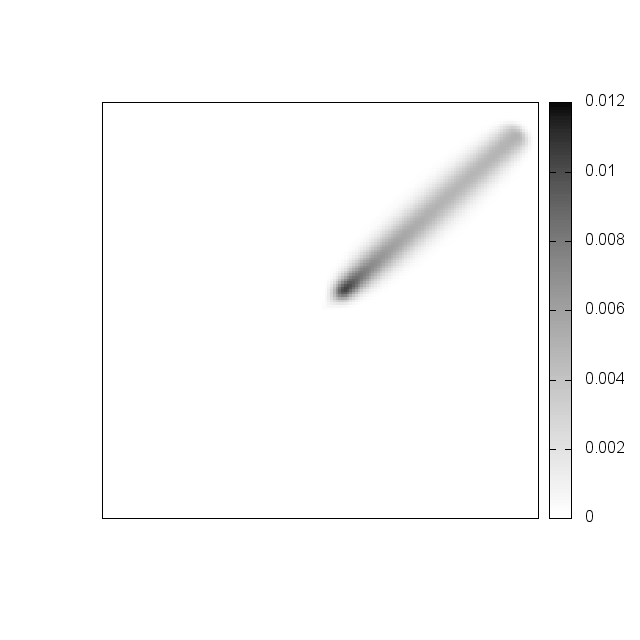}
$$
$$
\hspace{-0mm}\includegraphics[trim=70 70 10 50,clip,width=.33\textwidth]{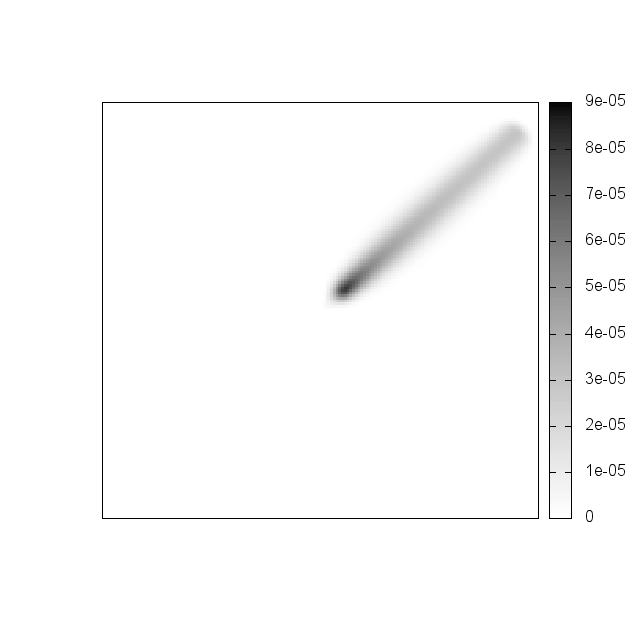}
\hspace{-0mm}\includegraphics[trim=70 70 10 50,clip,width=.33\textwidth]{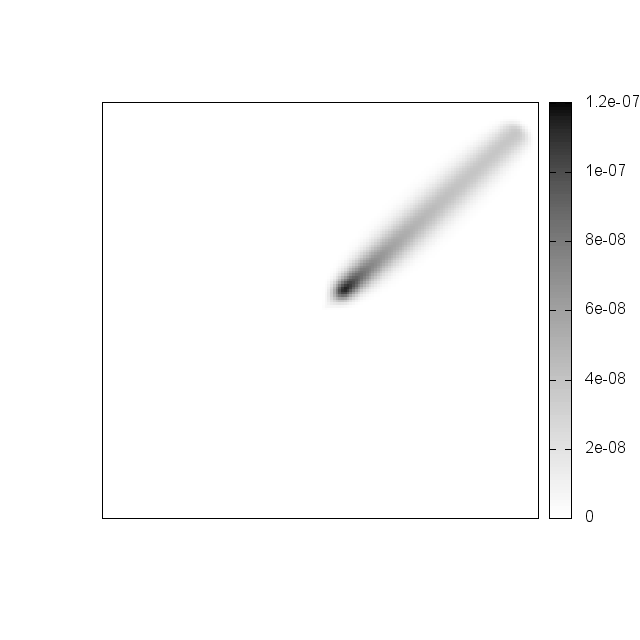}
\hspace{-0mm}\includegraphics[trim=70 70 10 50,clip,width=.33\textwidth]{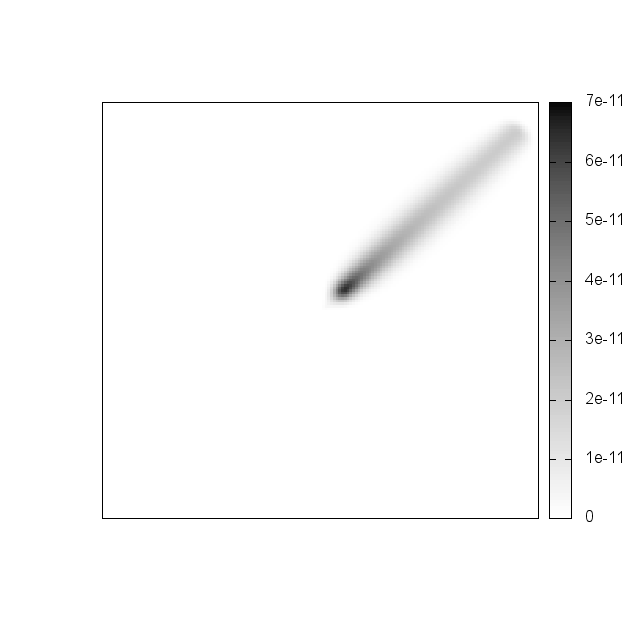}
$$
\caption{\small Returning ant density (left to right, top to bottom) from $t= 14 \min$ in increments of $14 \min$, up to the final time of 2.8 hours. Domain is $200\,\cm \times 200 \, \cm.$}
\label{FigU2}
\end{figure}

\begin{figure}[htbp]
$$
\hspace{-0mm}\includegraphics[trim=80 80 10 50,clip,width=.33\textwidth]{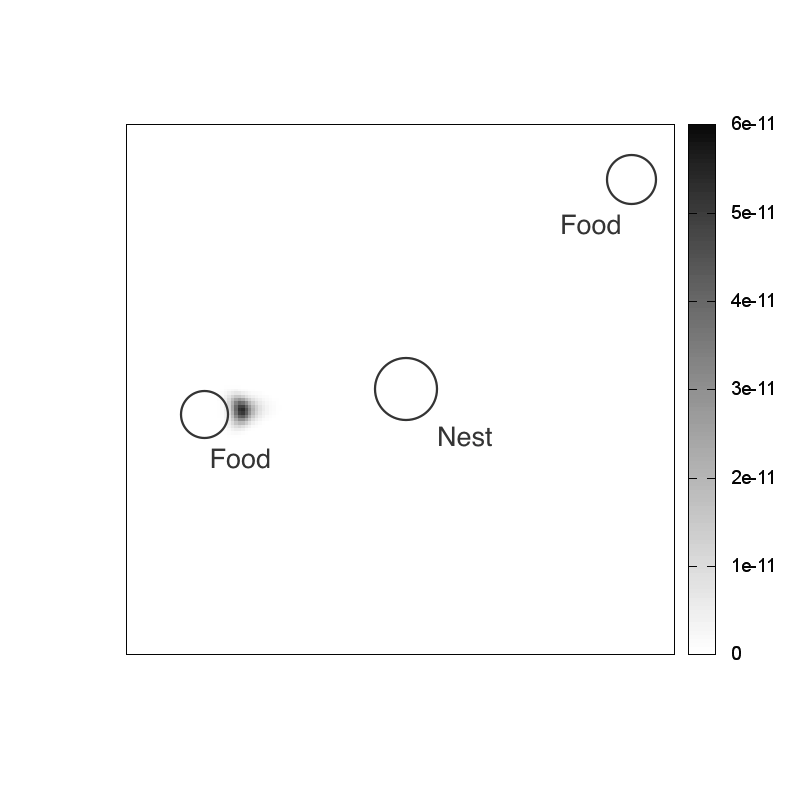}
\hspace{-0mm}\includegraphics[trim=70 70 10 50,clip,width=.33\textwidth]{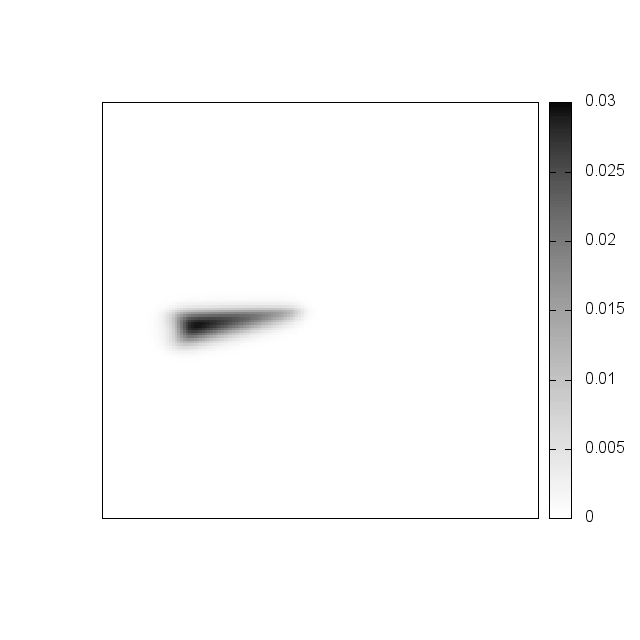}
\hspace{-0mm}\includegraphics[trim=70 70 10 50,clip,width=.33\textwidth]{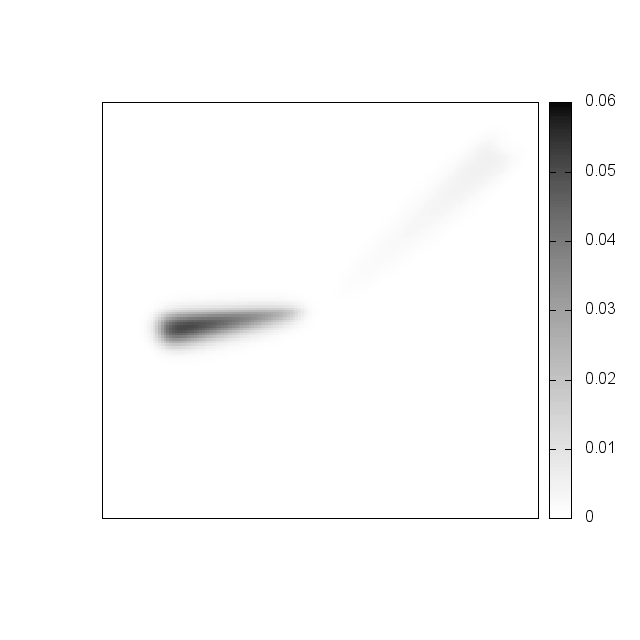}
$$
$$
\hspace{-0mm}\includegraphics[trim=70 70 10 50,clip,width=.33\textwidth]{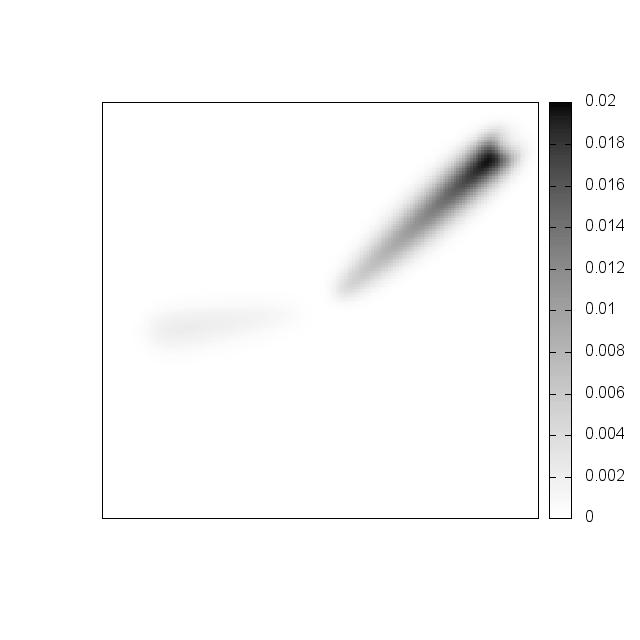}
\hspace{-0mm}\includegraphics[trim=70 70 10 50,clip,width=.33\textwidth]{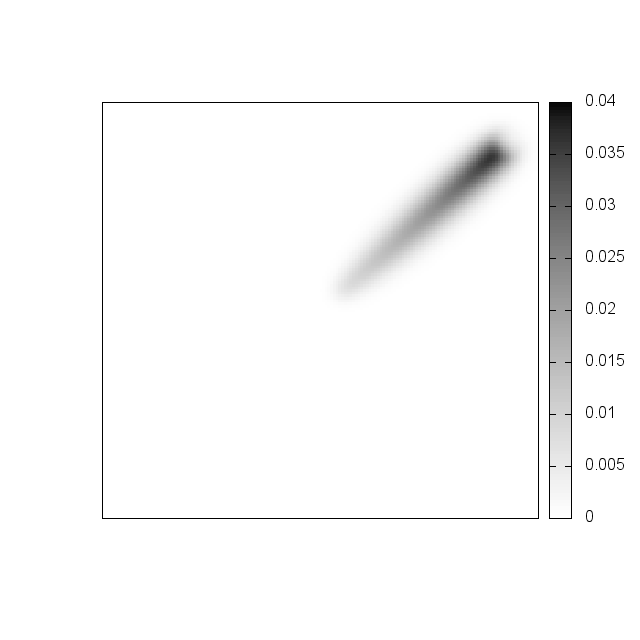}
\hspace{-0mm}\includegraphics[trim=70 70 10 50,clip,width=.33\textwidth]{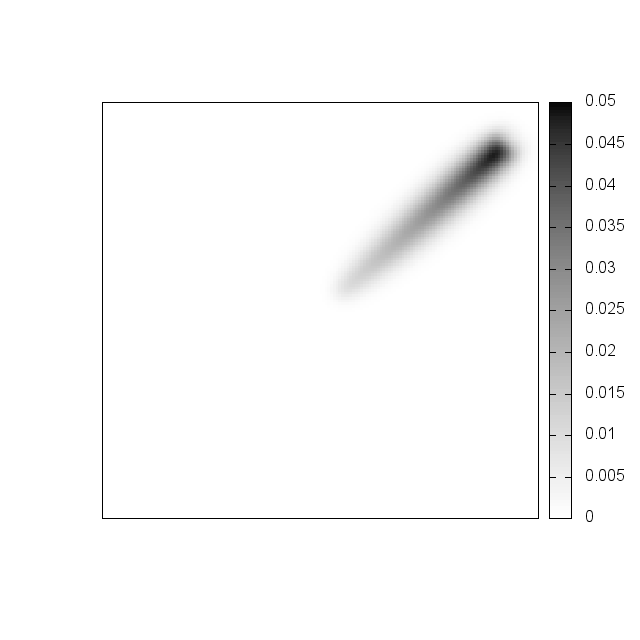}
$$
$$
\hspace{-0mm}\includegraphics[trim=70 70 10 50,clip,width=.33\textwidth]{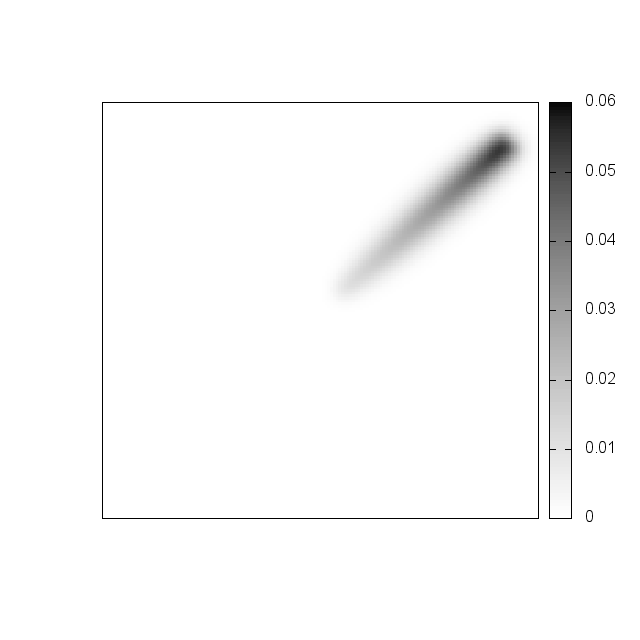}
\hspace{-0mm}\includegraphics[trim=70 70 10 50,clip,width=.33\textwidth]{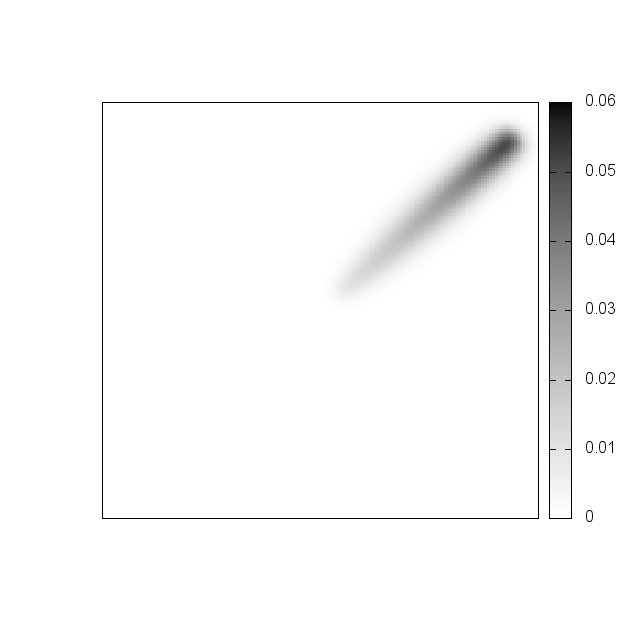}
\hspace{-0mm}\includegraphics[trim=70 70 10 50,clip,width=.33\textwidth]{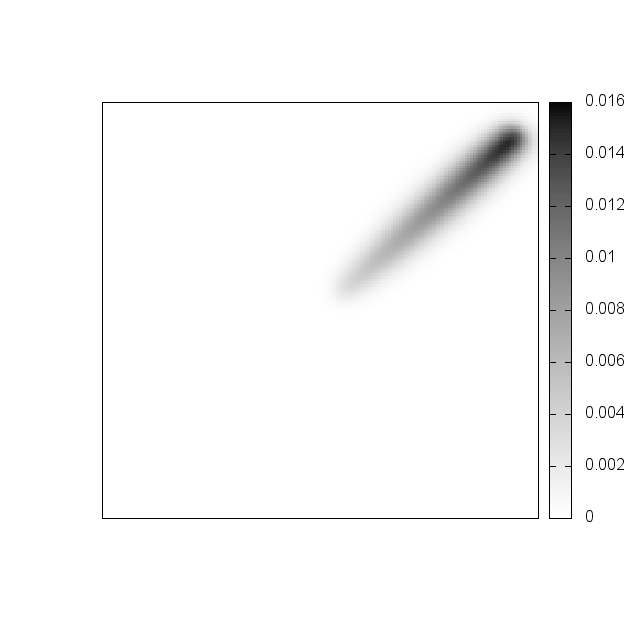}
$$
$$
\hspace{-0mm}\includegraphics[trim=70 70 10 50,clip,width=.33\textwidth]{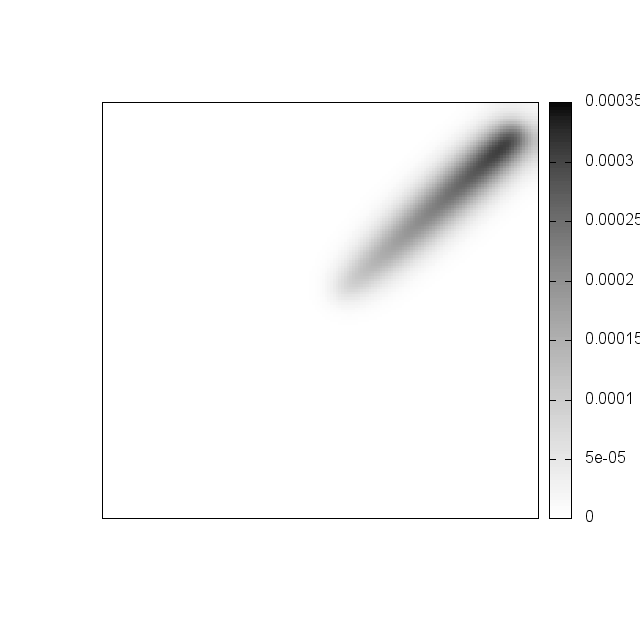}
\hspace{-0mm}\includegraphics[trim=70 70 0 50,clip,width=.33\textwidth]{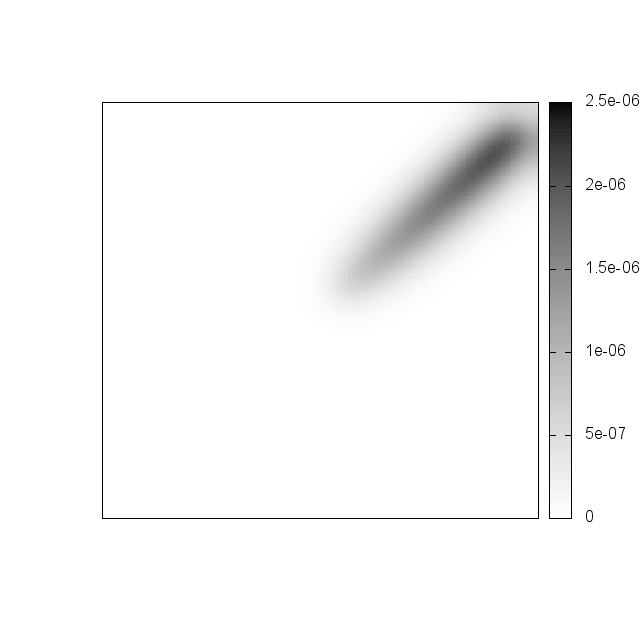}
\hspace{-0mm}\includegraphics[trim=70 70 0 50,clip,width=.33\textwidth]{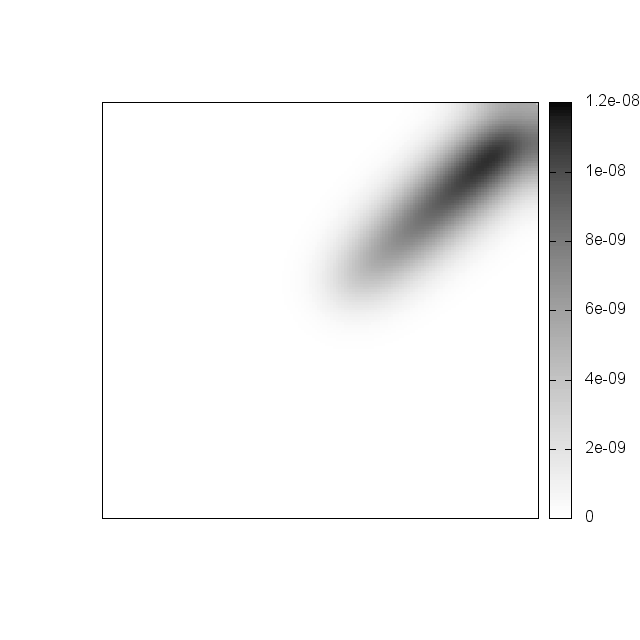}
$$
\caption{\small Pheromone concentration (left to right, top to bottom) from $t= 14 \min$ in increments of $14 \min$, up to the final time of 2.8 hours. Domain is $200\,\cm \times 200 \, \cm.$}
\label{FigPhero}
\end{figure}

\begin{figure}[htbp]
$$
\hspace{-0mm}\includegraphics[trim=0 0 0 0,clip,width=.9\textwidth]{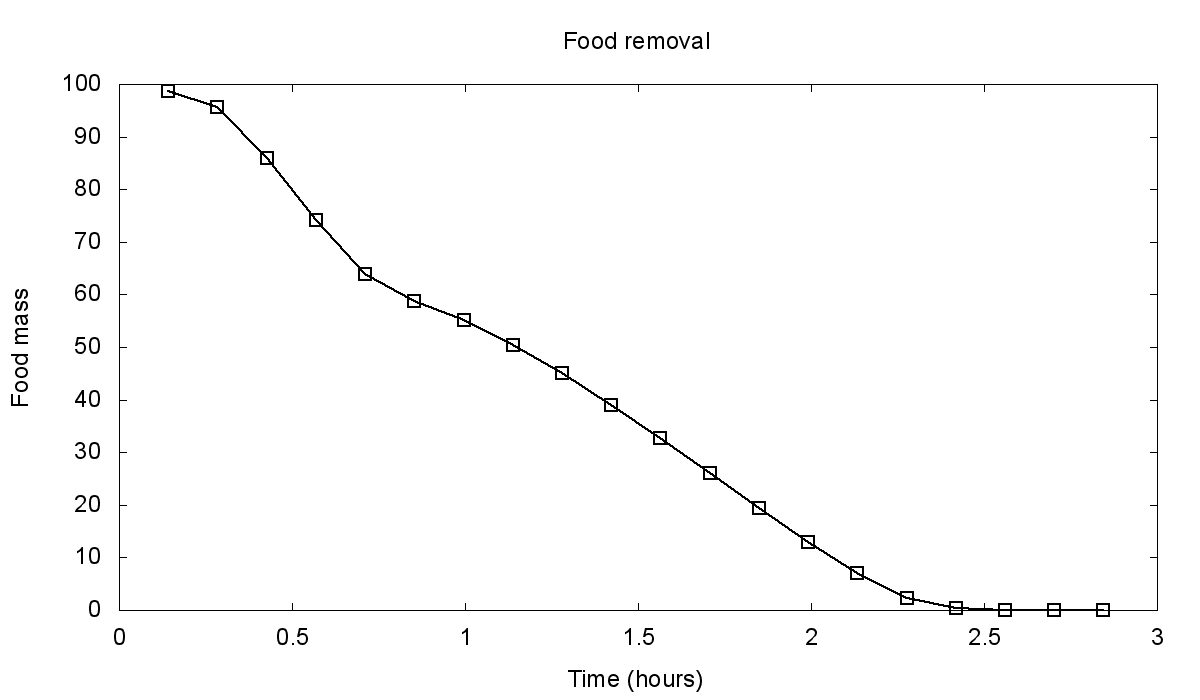}
$$
\caption{\small Evolution of food quantity. Horizontal axis in hours.}
\label{Fig029}
\end{figure}

In Figures \ref{FigU1}--\ref{Fig029}, we present the results of one typical simulation. We take a population of ants approaching the size of a young but established colony of, for instance, the fire ant \emph{S. invicta} or \emph{S. saevissima} \cite[p.160]{TheAnts}, \cite{FireAnts}. The data are collected in Table~\ref{T40}.

We present the ants with two different food sources. As discussed in the previous section, the lower value of food quantity on one of the sources reflects a lesser quality of the food source or simply a lower density of food. Even so, both food sources are eventually exhausted after about two and a half hours (see Fig.~\ref{Fig029}), which is consistent with the timescales reported in \cite{Wilson1962-1}. 

One can clearly observe the formation of trails in the foraging and in the returning ants, as well as the concentration of pheromone on that trail. Also, when the smaller food source is depleted, the evaporation of the pheromone results in a quick abandonment of the trail.

\begin{table}[htbp]
\caption{Parameters in system \eqref{2000}. See Table \ref{T20} for explanation. \label{T40}}
\begin{tabular}{ p{0.4\textwidth}|  p{0.5\textwidth}}
Parameter & Value \\ \hline
\hline&\\[-1.8ex] 
Total population & 75\,000 ants	  \\ [0.5ex]  
Area of physical domain $\Omega$ & $4\, \mathrm{m}^2$	  \\ [0.5ex]  
$\varepsilon$ & 0.5	  \\ [0.5ex]  
$D_v   $ & 0.1    \\ [0.5ex] 
$ D_w $& 0.1  \\ [0.5ex] 
$\chi_u$ &  50  \\ [0.5ex] 
$\lambda $ & 	70  \\ [0.5ex] 
$ c_0(x) $ &   $ \sim 12$  \\ [0.5ex] 
$\hat t  $ &  $102\, \seg$  \\ [0.5ex] 
$ \hat x $ &   $5.54 \,\cm$  \\ [0.5ex] 
Returning ant speed  & $1\, \mathrm{cm}/\mathrm{s}$	  \\ [0.5ex]  
\hline
\end{tabular}

\end{table}


\section{Food removal efficiency is linked to trail formation }
\label{Var}
In this section we discuss some consequences and possible experimental validations of our model. The main idea is that increased trail formation is correlated with food removal efficiency, which is in turn related to rather precise conditions on some of the coefficients of the system~\eqref{2000}.

In a simplified ecological setting such as the one presented in this work, the quantity which the ants strive to minimize (in an evolutionary sense) would be the amount of time taken to carry all the food to the nest.
So, in this section we investigate what is the influence on food removal efficiency of varying  parameters on which natural selection can act. 

Of particular interest is the question of whether different parameter combinations allow or prevent trail formation. To address this question, we focus our analysis on the pheromone degradation rate $\varepsilon$ and the chemotactic sensitivity of the foraging ants, $\chi_u$. Note that it would be interesting to  consider other pairs of parameters; however, these two parameters were chosen precisely because it is less clear, on an intuitive level, what effect their variation might have on the foraging dynamics.

\subsection{Parameter space exploration: conditions for trail formation}

With this in mind, we now present an exploration of the $(\varepsilon,\chi_u)-$parameter space, showing how trail formation in foraging ants is affected by the variation of these parameters. Recall that $\chi_u$ is the foraging ants' chemotactic sensitivity and $\varepsilon$ is the pheromone degradation rate.

In Figure \ref{FigTrail}, we present the density of foraging ants at time $t = 1.36$ hours, for various simulations using the same parameters as in Table~\ref{T20}, but with a smaller quantity of food. On the horizontal axis, $\varepsilon$ varies between the values 0.01, 0.1, 0.5, 1, 2.5 and 5, while on the vertical axis $\chi_u$ varies between  20, 40, 80, 160, 500 and 1000, from bottom to top. 

It is clear from Figure \ref{FigTrail} that certain parameter values are more conducive to trail formation than others. We can see, for instance, that trail formation is suppressed when the pheromone degradation rate $\varepsilon$ is large and the ants' sensitivity is low (bottom right of Figure~\ref{FigTrail}). Similarly, trails do not develop well when pheromone degradation is very fast and sensitivity very high (upper left part of Figure~\ref{FigTrail}).

\subsection{Trail formation and food removal efficiency}
As we pointed out, trail formation itself confers no advantage to the ants if it does not contribute to a more efficient food removal. We now show that trail forming behavior, as depicted in Figure~\ref{FigTrail}, is  correlated with increased food removal efficiency. To see this, we compute the evolution of the remaining food mass for each column and row of Figure~\ref{FigTrail}. 
Then, we determine for each row and column what are the parameter values for which food is more efficiently removed. The measure we use is the quantity of food remaining at a fixed time about 75\% of the total simulation time. The results are reported in Table~\ref{T30}. For each column of Figure~\ref{FigTrail}, that minimum is marked with a black star, while for each row the minimum is marked with a white star. These minima are also shown in Table~\ref{T30}.

Using this procedure, we can see from Table \ref{T30} and Figure \ref{FigTrail} that food removal is more efficient precisely when trail formation is more apparent. This strongly supports the notion that trail formation by foraging ants is a main contributing factor to a more efficient removal of the food, as is widely assumed.

Another conclusion suggested by the analysis of Table~\ref{T30} is that greater chemotactic sensitivity is not always advantageous; rather, increased sensitivity only translates into increased food removal efficiency when the pheromone degradation rate is sufficiently high. Conversely, a very long lived pheromone should be paired with a decreased sensitivity in order to yield efficient food removal. In other words, we can postulate a monotone dependence of the sensitivity with respect to the pheromone degradation rate.

\subsection{Suggestion of experimental work}

\begin{figure}[htbp]
$$
\hspace{-0mm}\includegraphics[trim=70 70 70 70,clip,width=.166\textwidth]{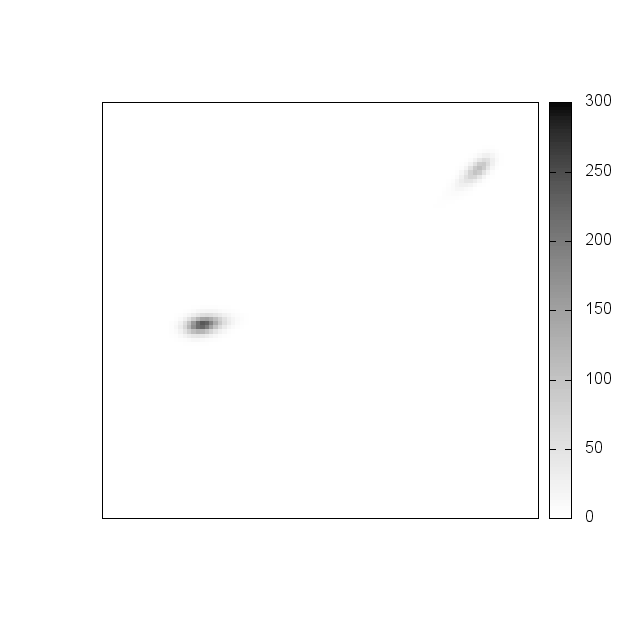}\hspace{-0mm}\includegraphics[trim=70 70 70 70,clip,width=.166\textwidth]{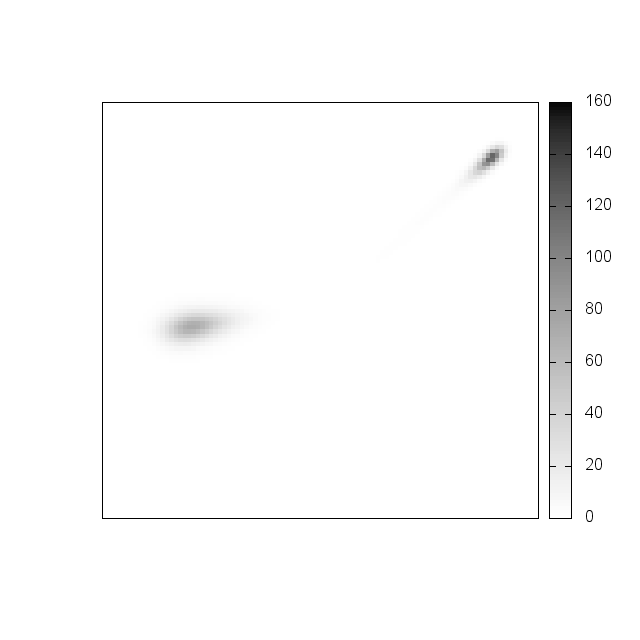}
\hspace{-0mm}\includegraphics[trim=70 70 70 70,clip,width=.166\textwidth]{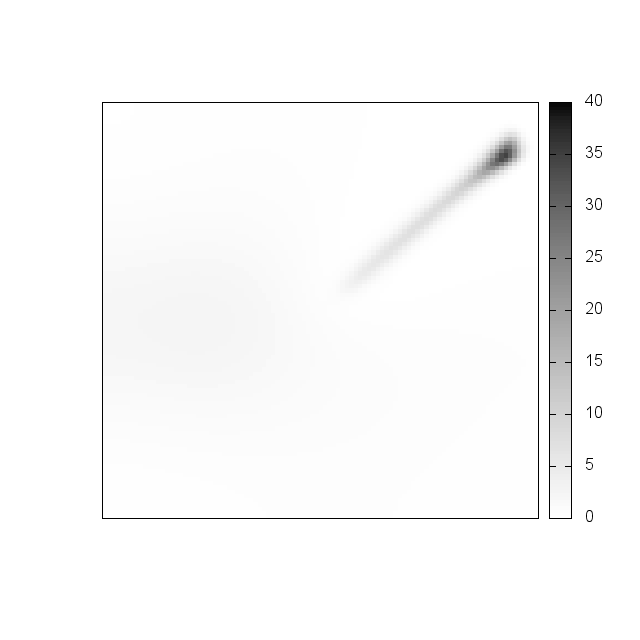}
{\hspace{-0mm}\includegraphics[trim=70 70 70 70,clip,width=.166\textwidth]{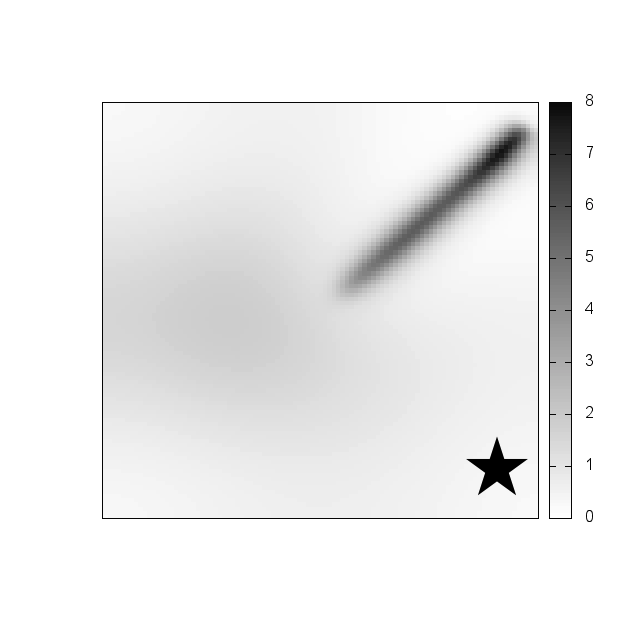}}
{\hspace{-0mm}\includegraphics[trim=70 70 70 70,clip,width=.166\textwidth]{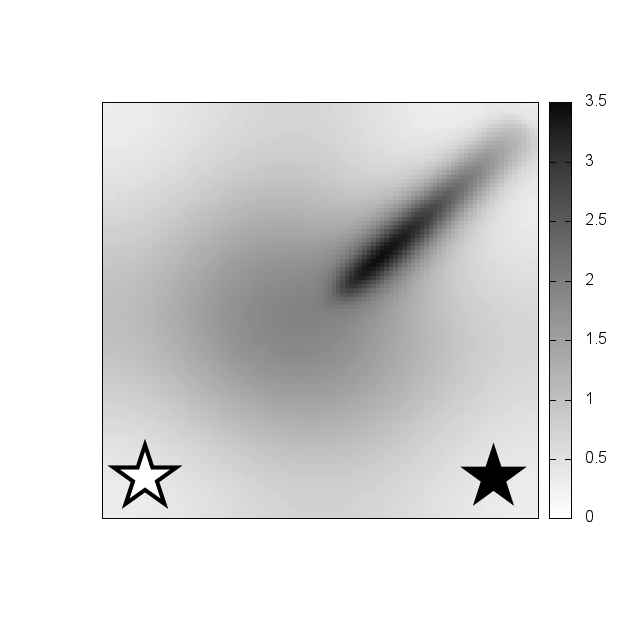}}
{\hspace{-0mm}\includegraphics[trim=70 70 70 70,clip,width=.166\textwidth]{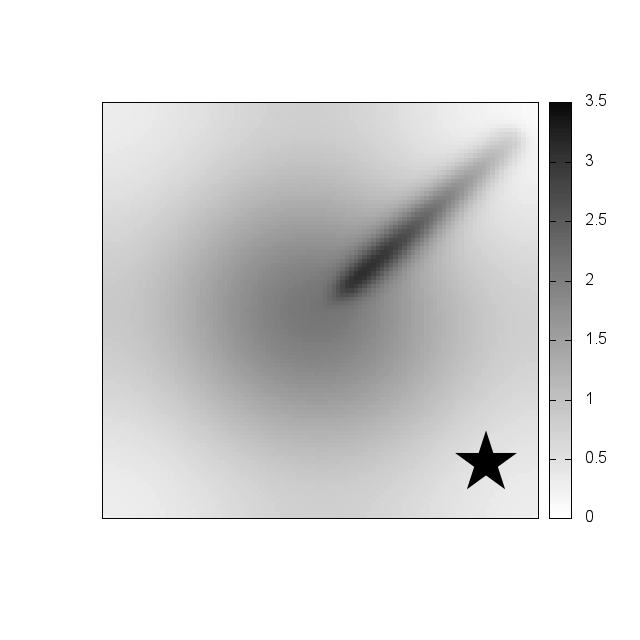}}
$$
$$
\hspace{-0mm}\includegraphics[trim=70 70 70 70,clip,width=.166\textwidth]{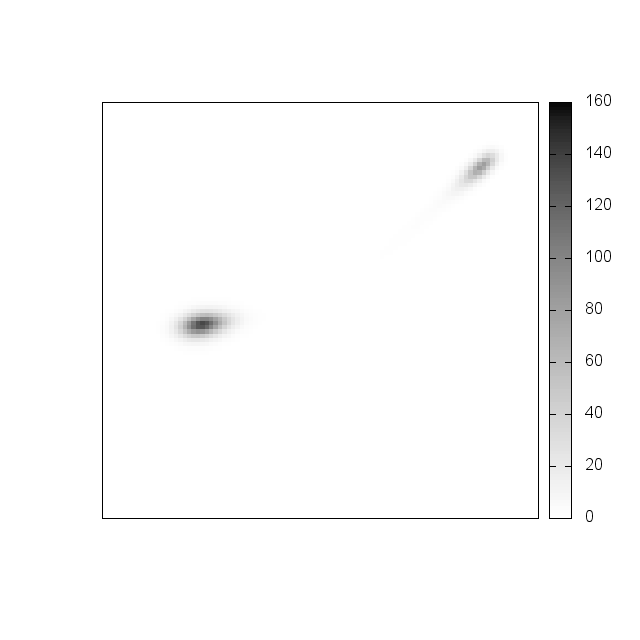}\hspace{-0mm}\includegraphics[trim=70 70 70 70,clip,width=.166\textwidth]{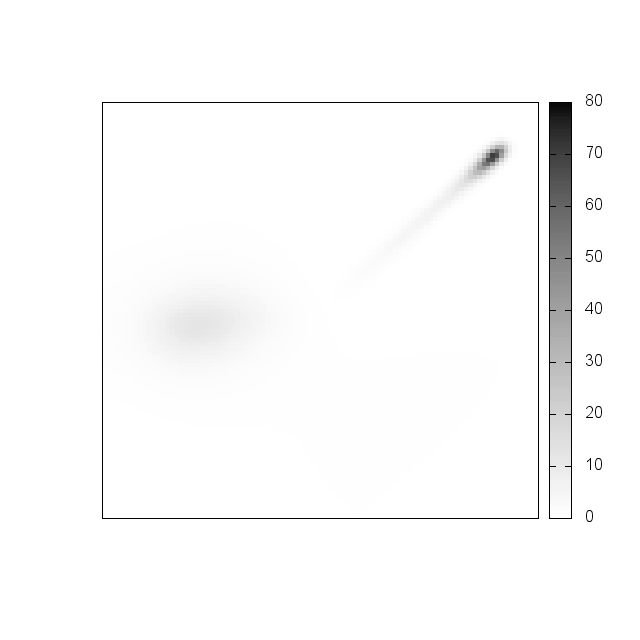}
\hspace{-0mm}\includegraphics[trim=70 70 70 70,clip,width=.166\textwidth]{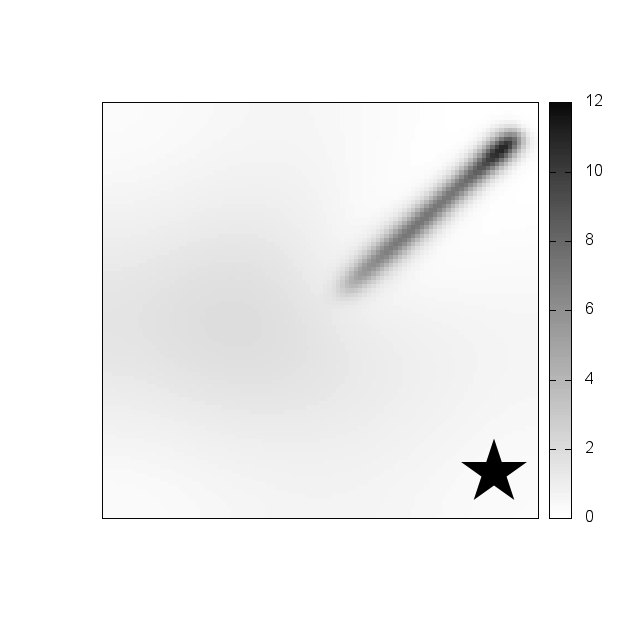}
{\hspace{-0mm}\includegraphics[trim=70 70 70 70,clip,width=.166\textwidth]{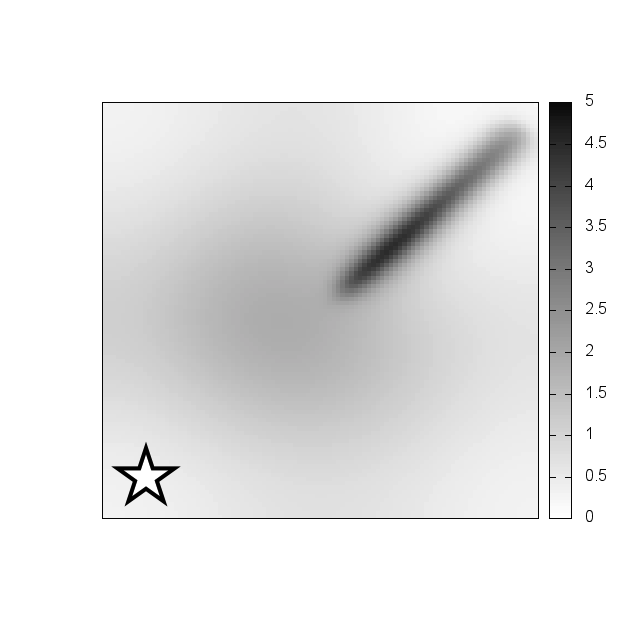}}
{\hspace{-0mm}\includegraphics[trim=70 70 70 70,clip,width=.166\textwidth]{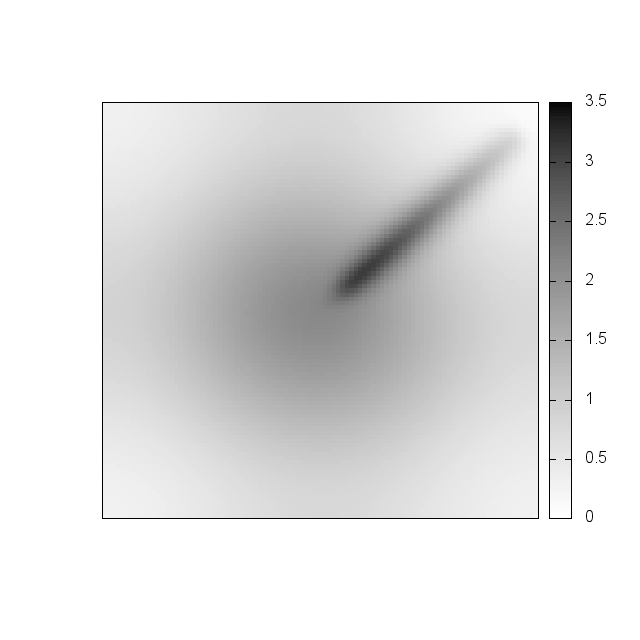}}
{\hspace{-0mm}\includegraphics[trim=70 70 70 70,clip,width=.166\textwidth]{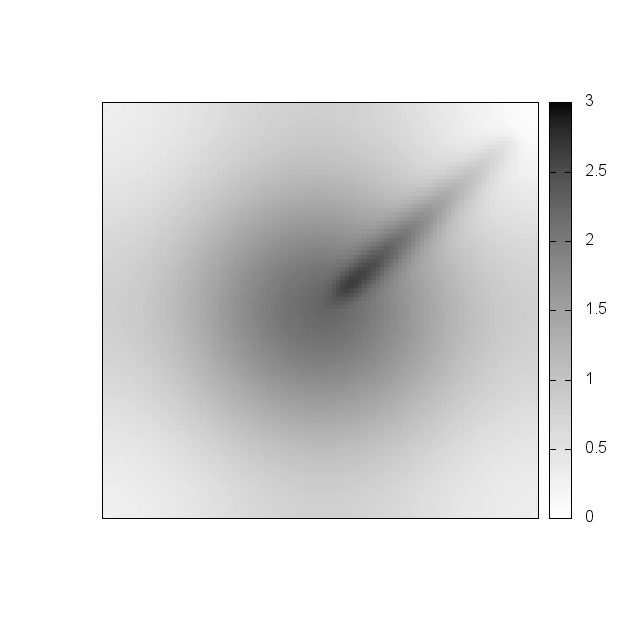}}
$$
$$
\hspace{-0mm}\includegraphics[trim=70 70 70 70,clip,width=.166\textwidth]{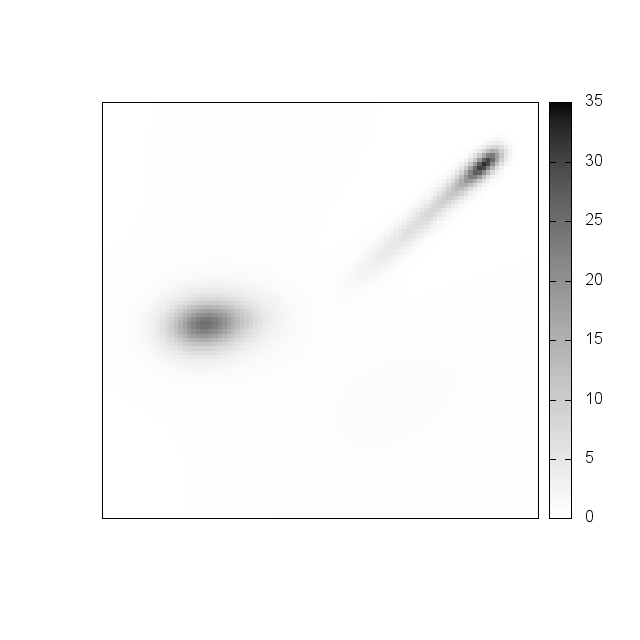}\hspace{-0mm}\includegraphics[trim=70 70 70 70,clip,width=.166\textwidth]{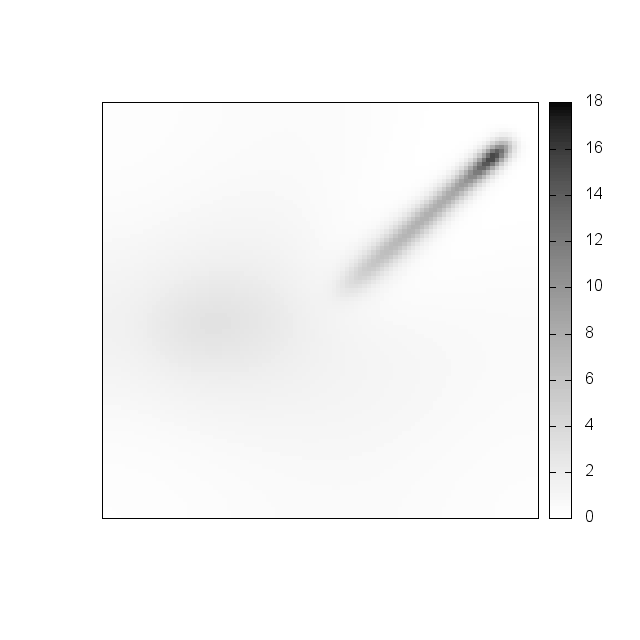}
{\hspace{-0mm}\includegraphics[trim=70 70 70 70,clip,width=.166\textwidth]{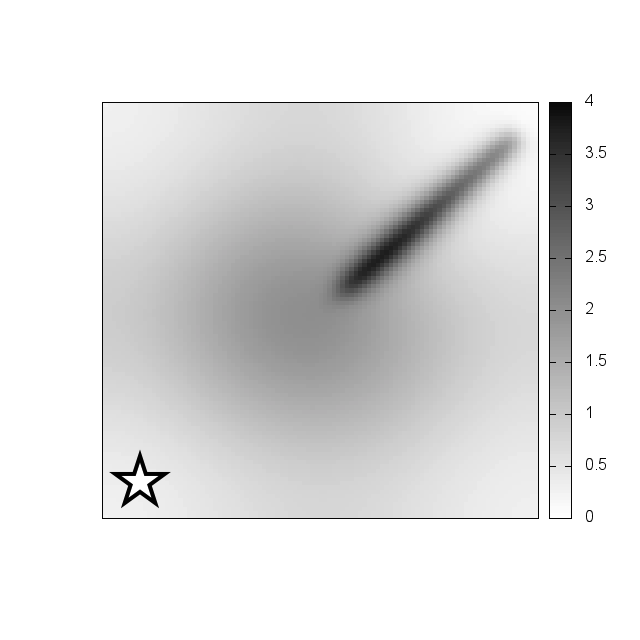}}
\hspace{-0mm}\includegraphics[trim=70 70 70 70,clip,width=.166\textwidth]{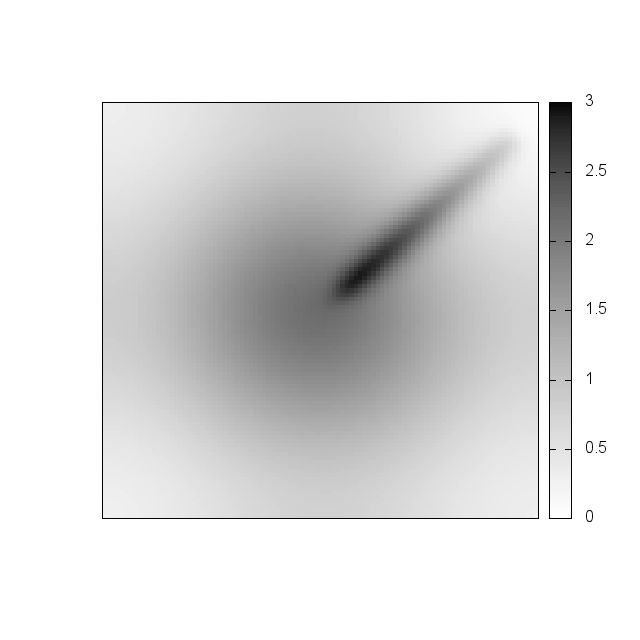}
\hspace{-0mm}\includegraphics[trim=70 70 70 70,clip,width=.166\textwidth]{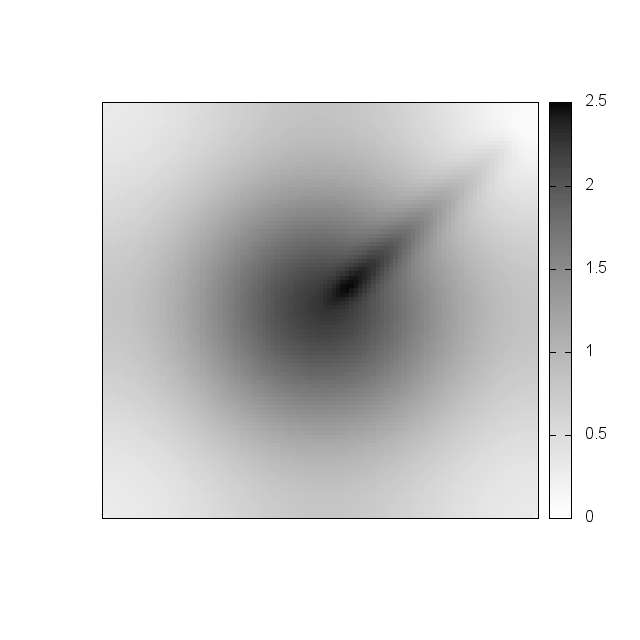}
\hspace{-0mm}\includegraphics[trim=70 70 70 70,clip,width=.166\textwidth]{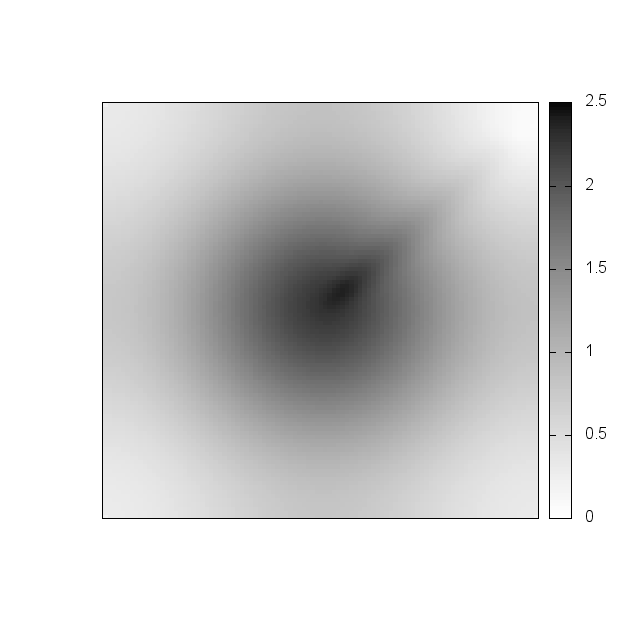}
$$
$$
\hspace{-0mm}\includegraphics[trim=70 70 70 70,clip,width=.166\textwidth]{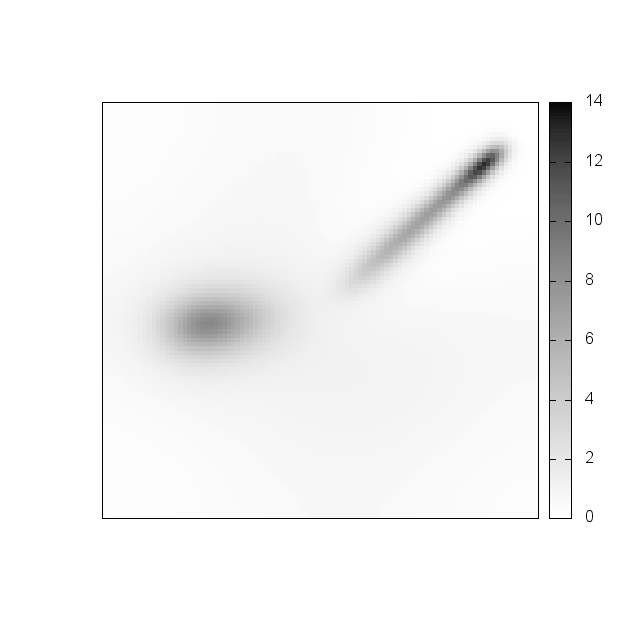}\hspace{-0mm}\includegraphics[trim=70 70 70 70,clip,width=.166\textwidth]{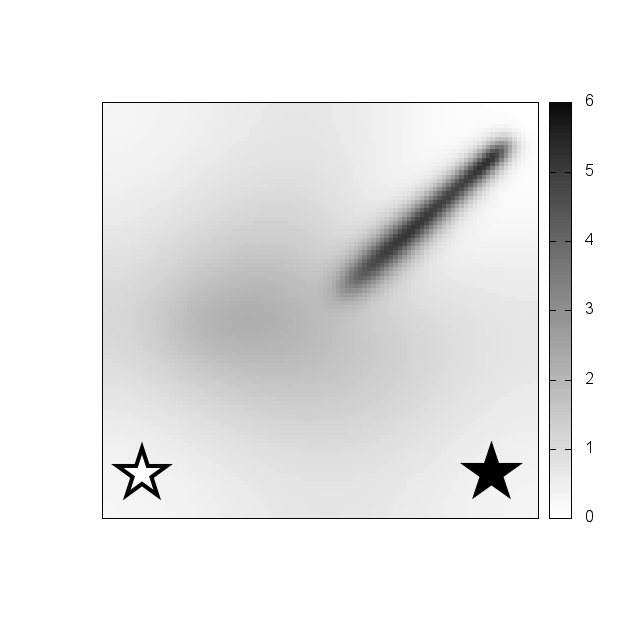}
\hspace{-0mm}\includegraphics[trim=70 70 70 70,clip,width=.166\textwidth]{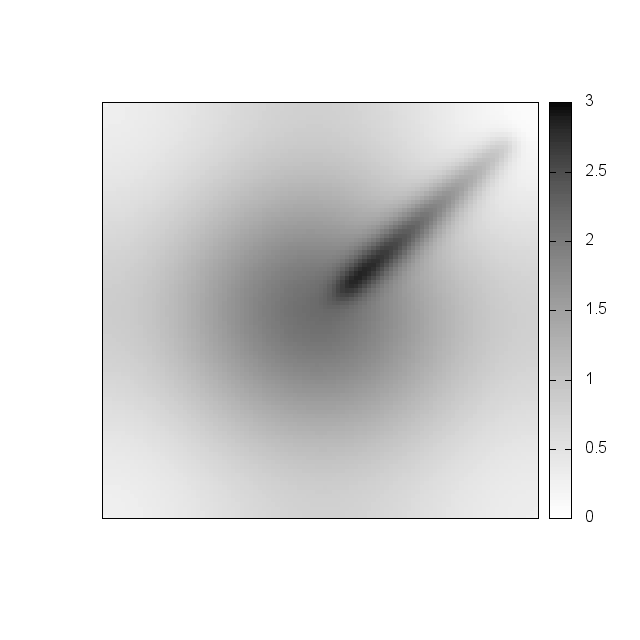}
\hspace{-0mm}\includegraphics[trim=70 70 70 70,clip,width=.166\textwidth]{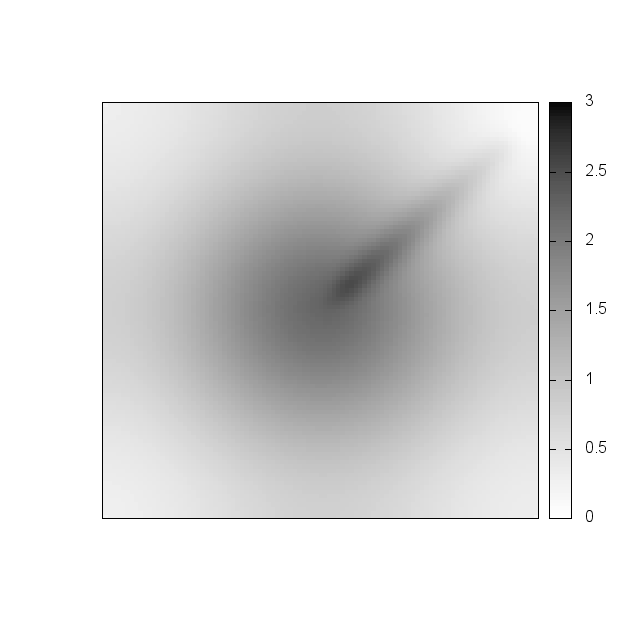}
\hspace{-0mm}\includegraphics[trim=70 70 70 70,clip,width=.166\textwidth]{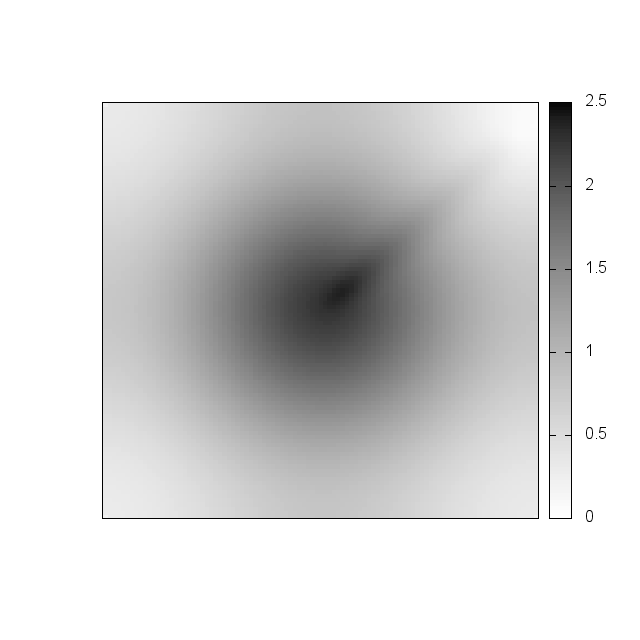}
\hspace{-0mm}\includegraphics[trim=70 70 70 70,clip,width=.166\textwidth]{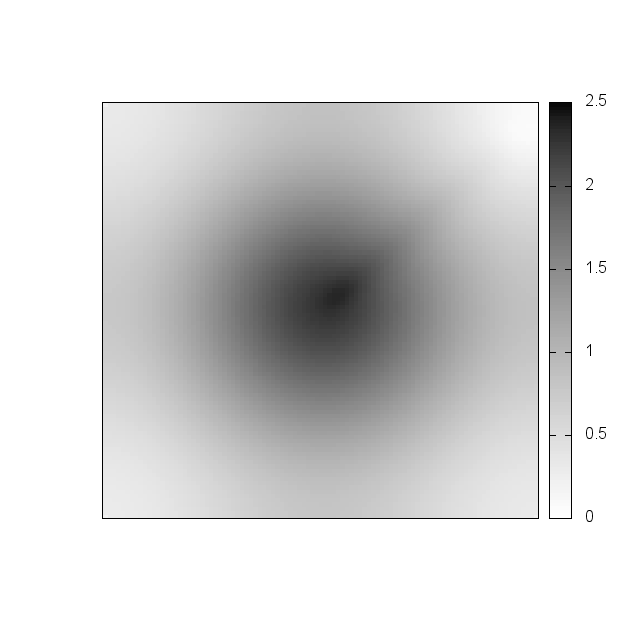}
$$
$$
\hspace{-0mm}\includegraphics[trim=70 70 70 70,clip,width=.166\textwidth]{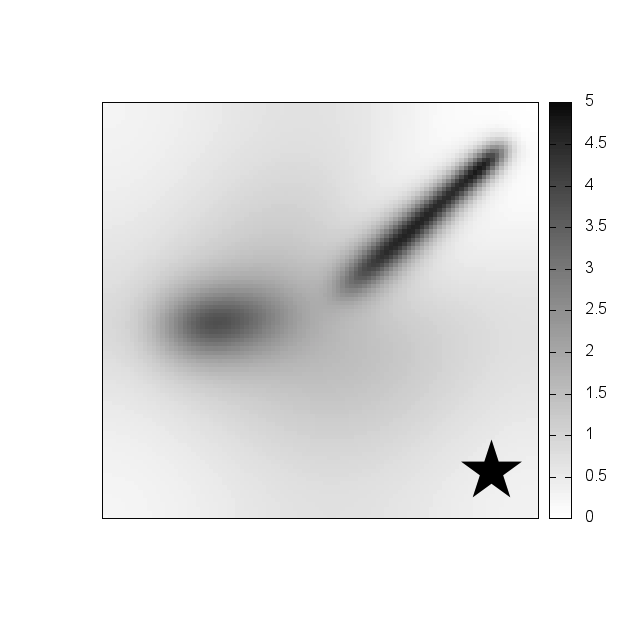}
{\hspace{-0mm}\includegraphics[trim=70 70 70 70,clip,width=.166\textwidth]{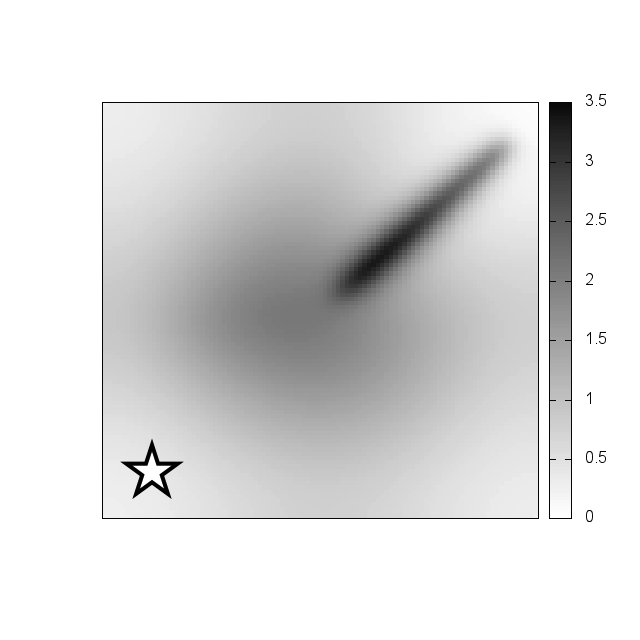}}
\hspace{-0mm}\includegraphics[trim=70 70 70 70,clip,width=.166\textwidth]{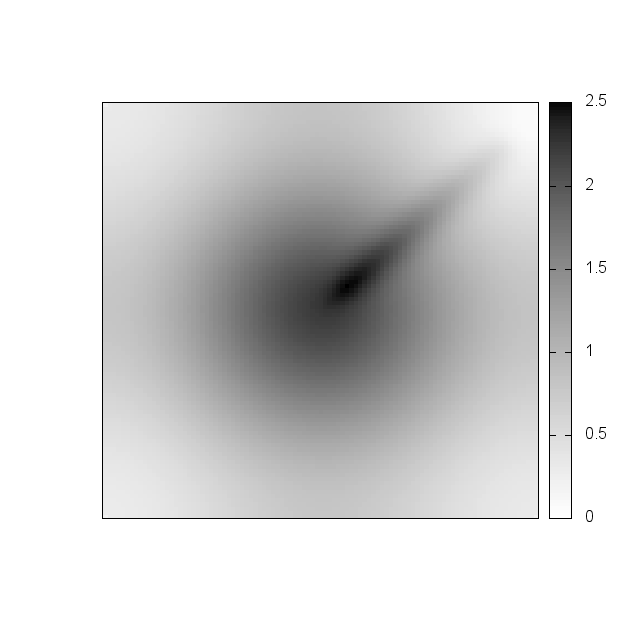}
\hspace{-0mm}\includegraphics[trim=70 70 70 70,clip,width=.166\textwidth]{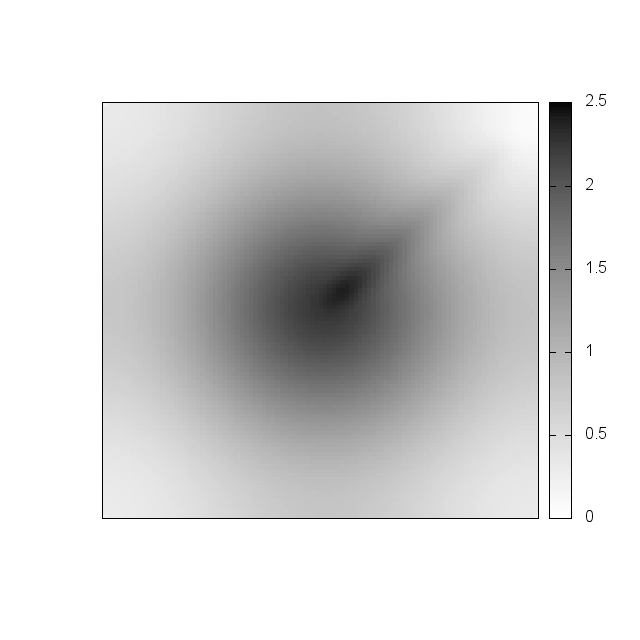}
\hspace{-0mm}\includegraphics[trim=70 70 70 70,clip,width=.166\textwidth]{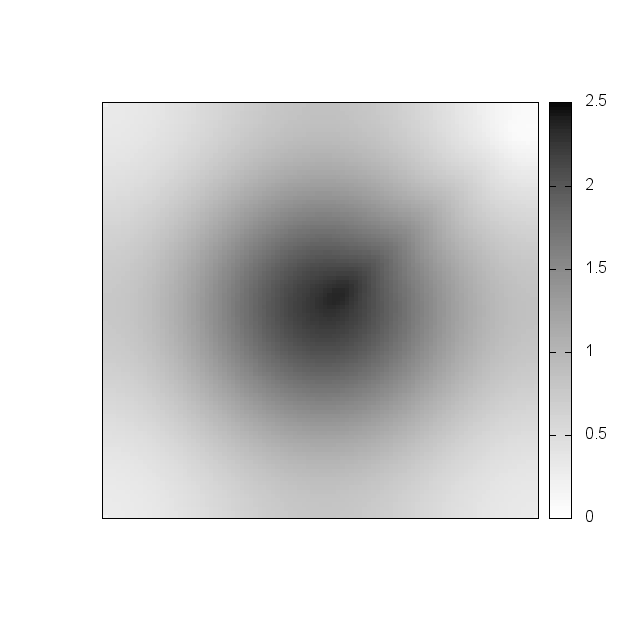}
\hspace{-0mm}\includegraphics[trim=70 70 70 70,clip,width=.166\textwidth]{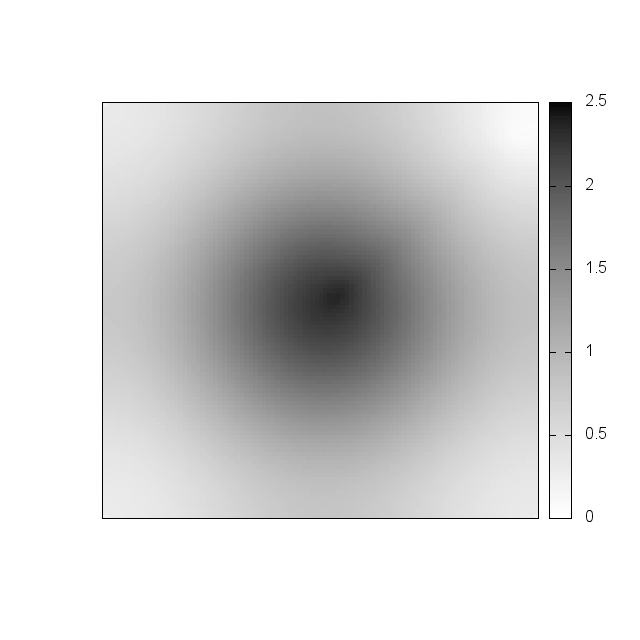}
$$
$$
{\hspace{-0mm}\includegraphics[trim=70 70 70 70,clip,width=.166\textwidth]{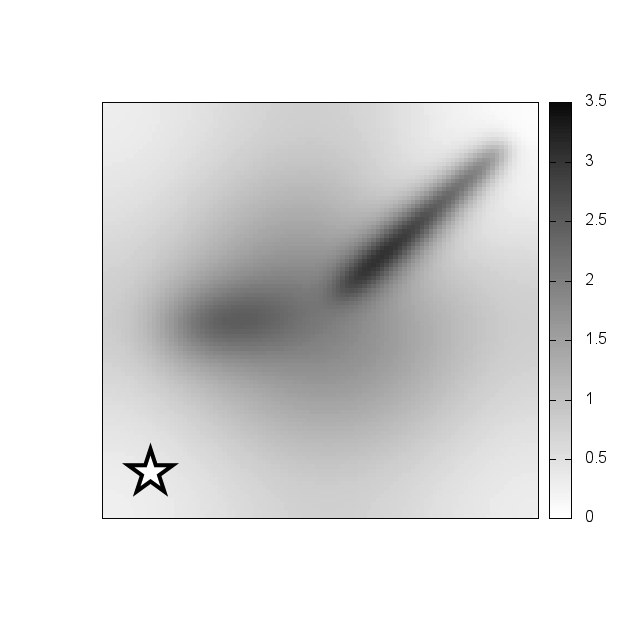}}\hspace{-0mm}\includegraphics[trim=70 70 70 70,clip,width=.166\textwidth]{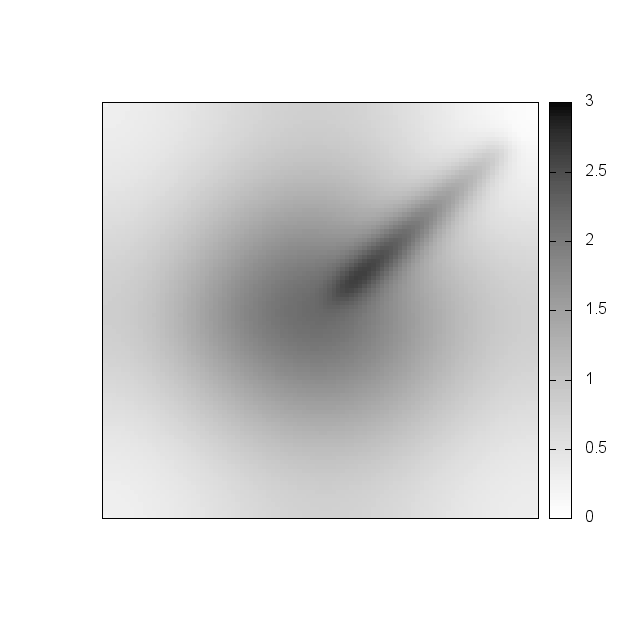}
\hspace{-0mm}\includegraphics[trim=70 70 70 70,clip,width=.166\textwidth]{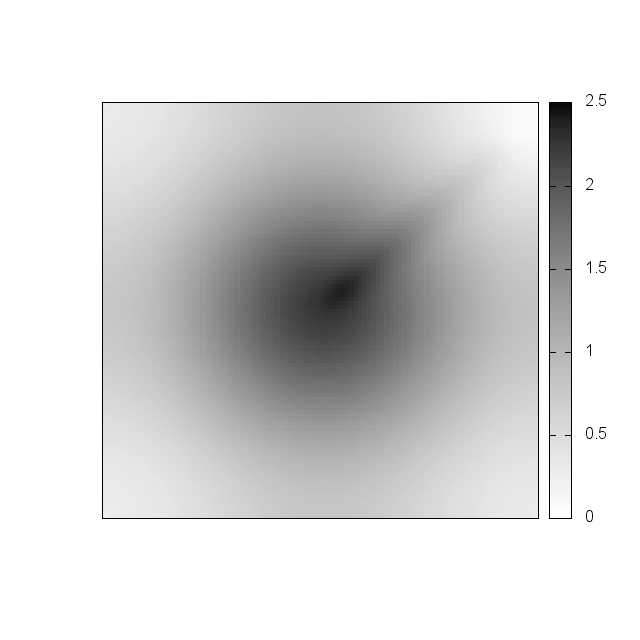}
\hspace{-0mm}\includegraphics[trim=70 70 70 70,clip,width=.166\textwidth]{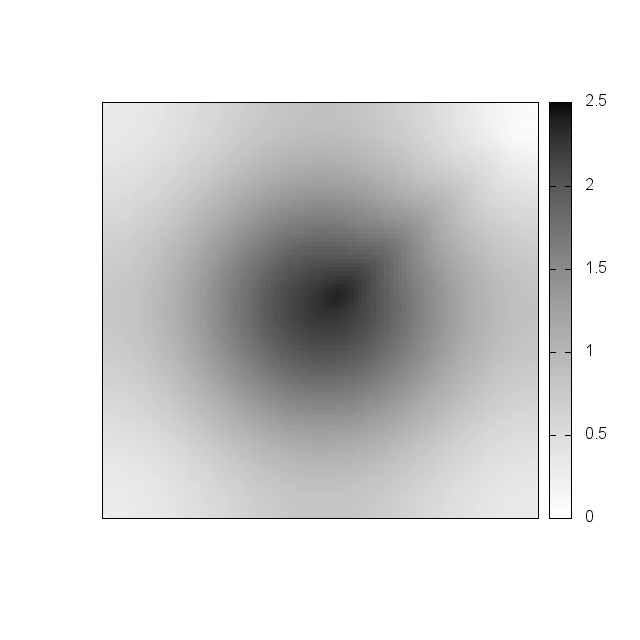}
\hspace{-0mm}\includegraphics[trim=70 70 70 70,clip,width=.166\textwidth]{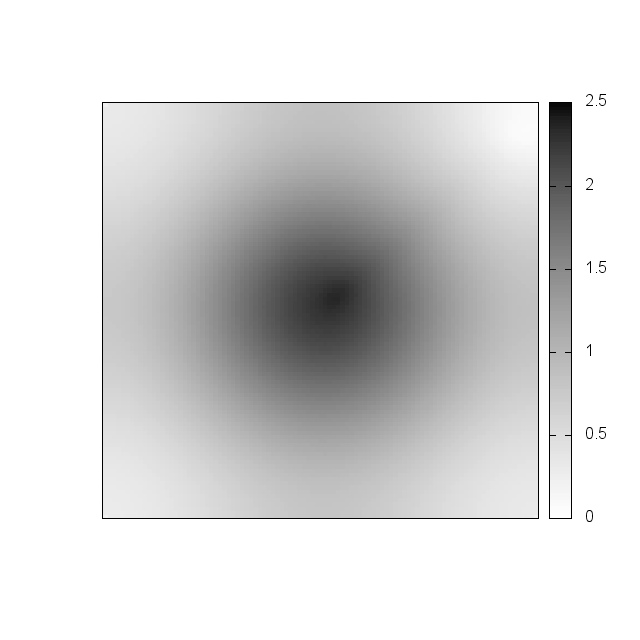}
\hspace{-0mm}\includegraphics[trim=70 70 70 70,clip,width=.166\textwidth]{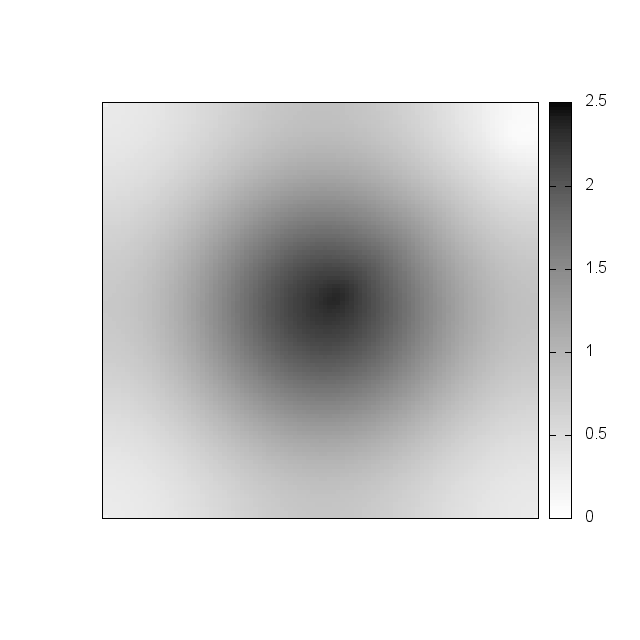}
$$
\caption{\small Conditions for trail formation in foraging ants. From left to right, $\varepsilon = 0.01, 0.1, 0.5, 1, 2.5, 5.$ Bottom to top, $\chi_u=  20, 40, 80, 160, 500, 1000.$ $\varepsilon$ is the pheromone degradation rate, $\chi_u$ is the foraging ants' chemotactic sensitivity (see Table \ref{T20}). A star $\star$ indicates simulations where, for fixed $\varepsilon$, the food removal efficiency is greatest. A white star indicates simulations where, for fixed $\chi_u$, the food removal efficiency is greatest (see Table~\ref{T30}).}
\label{FigTrail}
\end{figure}

\paragraph{Relation between pheromone degradation rate, chemotactic sensitivity and trail formation}
From Table \ref{T30} and Figure \ref{FigTrail} we can see that, in our simulations, there is a narrow region of the $(\varepsilon,\chi_u)$ parameter space leading to an optimal food removal efficiency (when one of the two parameters is fixed). Moreover, Figure~\ref{FigTrail} shows that in this region, trail laying is most apparent.

A possible experimental validation of our model would be to consider two closely related ant species which use different pheromones, where each pheromone has a different degradation rate. Our model predicts that, in that case, their sensitivities should lie in the narrow region of the $(\varepsilon,\chi_u)$ parameter space where food removal efficiency is greater. Given the appropriate parameters to plug into the equations \eqref{2000}, we would be able to predict the chemotactic sensitivity from the knowledge of the degradation rate alone, by constructing Table~\ref{T30}. 

To summarize, it may seem natural that chemotactic sensitivity varies with the pheromone degradation rate. What we have observed in our model is that not only is this a monotone dependence, but also that greater food removal efficiency occurs in a narrow region of the parameter space, which can, presumably, be known for a particular species of ant. Thus we would expect $\varepsilon$ and $\chi_u$ to be highly correlated in nature, at east for species which do not differ greatly in other respects.

Moreover, our simulations show that food removal efficiency is correlated with trail formation, reinforcing the view that trail formation is indeed one of the most important adaptations in ants.

\begin{table}[htpb]
\caption{Food removal efficiency as a function of pheromone evaporation rate and chemotactic sensitivity. A star ($\star$) indicates the minimum value for each column, and a diamond ($\diamond$) indicates the minimum value for each column.  \label{T30}}\medskip
\begin{tabular}{ |p{0.1\textwidth} ||  p{0.1\textwidth} |  p{0.1\textwidth}|  p{0.1\textwidth}|  p{0.1\textwidth}|  p{0.1\textwidth}|  p{0.1\textwidth} |}
\hline
$\chi_u\,\,\, \backslash\,\,\, \varepsilon$& \small{ $0.01$}  & \small{ $0.1$}  & \small{ $0.5$}  & \small{ $1$}  & \small{$2.5$} & \small{ $ 5$}     \\ \hline
\hline&\\[-1.8ex] 
\small{$ 1000$} &16.58 & 10.51 & 3.67 & 1.35 \hfil $\star$ & 1.14 \hfil $ \diamond\star$  & 3.16\hfil $\star$   \\ [0.5ex]  
\small{$ 500$} 	& 11.95 & 7.99 & 2.56\hfil $\star$ & 1.69 \hfil $\diamond$& 3.34 & 4.93    \\ [0.5ex]  
\small{$ 160$} 	& 7.08 & 4.7 & 3.22 \hfil $\diamond$ & 4.26 & 5.61 & 6.14  \\ [0.5ex]  
\small{$ 80$}	& 5.18 & 4.02 \hfil $\diamond\star$ & 4.59 & 5.45  & 6.16 & 6.41  \\ [0.5ex]  
\small{$ 40$}	& 4.53\hfil $\star$ & 4.5 \hfil $\diamond$ & 5.6 & 6.07 & 6.42 & 6.55   \\ [0.5ex]  
\small{$ 20$}	& 5.02 \hfil $\diamond$ & 5.39 & 6.15 & 6.38 & 6.55 & 6.61 \\ [0.5ex]  
\hline
\end{tabular}
\end{table}

\paragraph{Orientation along trails}

From our review of the myrmecological literature, one outstanding open problem appears to be the question of the choice of direction when encountering a trail (this was discussed in detail in Section~\ref{Mod}). The present work suggest that the ability of ants to measure differences in pheromone concentration along the length of the trail, and not only across the trail, may play an important role in orienting individuals in the correct direction when encountering a trail. This lengthwise gradient can be created by a gradual diminishing of pheromone deposition by returning ants as they approach the nest. Indeed, although other mechanisms to solve this orientation problem have been found, and others can be envisaged, our results suggest that gradual diminishing of pheromone deposition can contribute to allow ants to find the correct orientation when encountering a pheromone trail, especially in the case of relatively short trails (on the order of one to a few meters) as the ones considered here. Thus it could be of interest to carry out more detailed experiments to ascertain whether lengthwise variation in pheromone concentration along the trail plays any role in individual orientation along the trail.

\section{Conclusions and future work}
\label{Sec040}

\subsection{General conclusions}
We have presented a mathematical model of ant foraging using a system of PDEs in the mathematical framework of chemotaxis. We have shown numerically that this system can exhibit spontaneous trail formation in the presence of food sources. The fact that trails are formed by the returning ants is built into our modeling assumptions and as such is not surprising; but we have shown, in addition, that trail formation occurs also in foraging ants, which is not explicitly stated in the model. 

Moreover, we have shown through parameter exploration that trail formation in foraging ants is sensitive to parameter values and, more importantly, is correlated with increased food removal efficiency. This allowed us to postulate that chemotactic sensitivity and pheromone degradation rates should also be correlated in a way that can be made precise for particular species of ants, and consequently tested by actual experiments. 

Our model allows for the simulation of a whole cycle of food foraging, from ants emerging from a nest onto a foraging ground where food is placed, discovering the food and returning to the nest laying pheromones. Recruitment then takes place, with foraging ants being attracted to the pheromones laid previously by the returning ants. A feedback loop ensues, as more ants reach the food source and return to the nest laying pheromones. Finally, when the food source is exhausted, the trail fades away due to the natural evaporation and diffusion of the pheromone.

\subsection{Limitations and future work}
\label{Sec9999}

Naturally, it would be very difficult for the model presented here to provide a comprehensive description of the extremely complex dynamics taking place during the whole foraging cycle of ants. One limitation is that we do not attempt to model in detail what happens near the nest and near the food site. In our model, only chemical signals intervene in the ants' behavior, whereas it is well known that individuals rely on a variety of other sensory and communication input to adjust their behavior. 

Another clear limitation is the functional form of the last equation of \eqref{500}, governing the removal of food. Intuitively, it would be more reasonable that each ant reaching the food would consume or carry away a fixed food quantity, rather than a proportion of the available food. As observed in an Section~\ref{Mod}, this choice was made primarily to ensure mathematical simplicity and facilitate the nondimensionalisation procedure. The model can, however, be easily adapted to more realistic food removal models.

Another possible improvement is in the chemotactic transport term for the $u$-equation 
in the system~\eqref{2000}. Indeed, its present form is not entirely realistic, since the (scalar) ant speed, given by $|\chi_u \nabla v|$ can have large variations, and in particular can theoretically attain very high values. But the actual movement of ants along trails shows little correlation between ant speed and ant density, see \cite{John,Johnson2006}. This is coherent with the informal observation that even on dense trails, ants move along at speeds similar to isolated ants, in contrast to, say, vehicles on a road. 
Therefore, a different modeling of the transport term could be envisaged to account for this lack of the so-called jammed phase in the flux. 

Related to this, is the question of the possible unlimited growth of ant density on trails. In the chemotactic literature, it is usual to introduce a term of the form $u_{\mathrm{max}}-u$ in the transport term, which serves as a limiter for the density: when the density reaches the value $u_{\mathrm{max}}$, individuals do not move and so density will not increase further. This is called the jammed phase and is usual in the modeling of vehicular or pedestrian traffic. It has been studied, for instance, in \cite{HillenPainter2001}. We find that such a mechanism is not realistic in the case of ants, since as we pointed out before, no jammed phase is observed \cite{John}. Other modeling strategies could be deployed to account for a limitation in density. Remark that the forthcoming results in \cite{AAG} actually give a uniform bound for all time on the density of ants by some constant $C$. However, $C$ comes from the mathematical analysis and so is not directly related to physical constraints on the ants' density.


Additionally, it would be more realistic to introduce a transport term in the foraging ant equation  expressing ants' tendency to stay near the nest, as has been done in other models of animal movement \cite{Lewis93,Lewis97}. This may prevent unrealistic situations such as, on the whole $\RR^2$, with no food present, the density of foraging ants decreasing to zero everywhere due to the diffusive effect. We omit such a term for the sake of keeping the model as simple as possible.

Finally, as far as we could gather, no experiment has been done to discover exactly what ``diffusion law'' is obeyed by ants. In this work we use the usual Fickian diffusion since it is the simplest and yields good results in a first approximation. However, it is well known that animal movement may be better approximated by other models of diffusion, usually nonlinear models (see for instance the book \cite{Okubo}). Thus more realistic diffusions could be incorporated in the model.

%

\appendix
\section{Nondimensionalisation}
\label{nondim}

Here, we will describe the nondimensionalisation procedure used to reduce the system \eqref{500} to the system \eqref{2000}. We will use the notation
\[
\aligned
t = \hat t t^*, \quad x = \hat x x^*, \quad u = \hat u u^*, \quad v = \hat v v^*, 
\endaligned
\]
and so on, for the changes of variable involved in nondimensionalisation. Here, $t$ (say) represents the old variable, $t^*$ the new, nondimensional variable, and $\hat t$ the (dimensional) new scale. Thus, for instance, $u^*$ is the (new) function defined through $\hat u \,u^*(t^*,x^*) =  u(\hat t t^*, \hat x x^*)$.

As is standard procedure, we will formulate the system \eqref{500} under the new unknowns $t^*,x^*, u^*,\dots,$ and finally remove the ${}^*$ for convenience. Nondimensionalisation therefore consists in a judicious choice of $\hat t, \hat x,$ and so on.

Our nondimensionalisation procedure will differ from the usual procedure used often in chemotaxis \cite[p.191]{HillenPainter2009}, in which the time scale is associated with the decay time (or half-life) of the attracting chemical. That would not be convenient in our case since this is a highly variable (and usually very short) quantity \cite{Bossert1963} and presents problems from the modeling point of view (as described in Section~\ref{phero} above).
Thus we set

\be
\label{550}
\aligned
\hat u = \hat w = \uref,
\endaligned
\ee
so that the density of foraging ants is measured as a proportion to this homogenous steady state of a uniformly distributed population.

Since the main objective of foraging is the efficient removal of food sources, it seems natural to consider a time scale tied to the rate of food removal by ants. Considering the last equation in \eqref{500} leads to the choice
\be
\label{600}
\aligned
\hat t = \frac1{\gamma \uref}.
\endaligned
\ee
The physical meaning of $\hat t$ is seen by observing that if the half-life of a food source is measured in the units of system \eqref{500} as, say, $t_0$ (from our modeling assumptions, i.e., the fourth equation in \eqref{500}, $t_0$ does not depend on the initial food quantity), then assuming constant in space food concentration and foraging ant density of $\uref$, the scale will be $\hat t = t_0 /\ln 2$ (in the units of $t_0$). 

Thus, the last equation of \eqref{500} becomes simply
\[
\aligned
\del_{t^*} c^*  = - u^*c^*.
\endaligned
\]

Next, choosing 
$$
\hat x = \sqrt{ \frac{\alpha_u}{\gamma \uref}}, \qquad \hat v = \frac\mu\gamma
$$
yields
\[
\aligned
\del_{t^*}v^* = P(x^*) w^* - \varepsilon v^* + D_v \Delta v^*,
\endaligned
\]
with
\[
\aligned
D_v = \frac{\alpha_v}{\alpha_u},\qquad \varepsilon = \frac\delta{\gamma \uref},\qquad P(x^*) = \overline P(\hat x x^*).
\endaligned
\]
Proceeding similarly with
\[
D_w = \frac{\alpha_w}{\alpha_u},  \qquad \chi_u = \frac{\beta_u\mu}{\alpha_u\gamma}, 
\qquad \lambda = \frac{\lambda_2}{\gamma \uref}, 
\]
\[
\hat c =\frac{\gamma \uref}{\lambda_1}, 
\qquad \hat a =\frac{\alpha_u}{\beta_w},
\]
\[
N(x^*) = \overline N(\hat x x^*), \qquad M(x^*) = \frac{\hat t}{\uref} \overline M(\hat x x^*),
\]
gives for the remaining two equations
\[
\aligned
&\del_{t^*} u^* -  \Delta u^* + \dive\big( u^*  \chi_u\nabla v^*  \big) = -  u^*c^* + \lambda w^* N(x^*) + M(t^*) N(x^*)
\\
&\del_{t^*} w^* - D_w \Delta w^* + \dive\big( w^*  { \nabla a}^*   \big) =  u^*c^* - \lambda w^* N(x^*)
\endaligned
\]

As is standard practice, we omit the ${}*$ from all the variables. 
Collecting the previous equations we obtain the nondimensional system \eqref{2000}.

\section{Details of the numerical scheme}
\label{numerical}

In this appendix we describe the numerical method used to integrate the system \eqref{2000}--\eqref{2600}.
For the spatial discretization, we divide the 2-d computational domain into rectangular cells of sides $dx$, $dy$. We use conservative schemes throughout: the Laplacian terms are discretized using standard centered differences, while the advection terms are discretized using a conservative first-order upwind scheme (see \cite{GR,LeVeque}).

The spatial discretization allows us to reduce the system \eqref{2000}--\eqref{2600} to a system of ordinary differential equations, which we integrate in time using a fourth-order Runge--Kutta method for systems (see \cite[p.331]{Burden}). We also used an explicit Euler scheme and compared the results of both methods to validate the results.

The upwind method used is known for its stability in dealing with advection terms. However, being a first order scheme, it introduces some numerical viscosity which tends to smear out results. It would thus be interesting, for future works, to implement a more accurate method, such as the high resolution methods found in \cite{LeVeque}.

\subsection*{Acknowledgements}
The author wishes to thank the referees for insightful suggestions which contributed greatly to the improvement of the paper. The author was partially supported by FAPERJ grant no. APQ1 - 111.400/2014  and CNPq grant no. Universal - 442960/2014-0.



%


\begin{thebibliography}{10} 


\bibitem{Amorim}
Amorim, P.,
A continuous model of ant foraging with pheromones and trail formation.
To appear in Proceedings of XXXV CNMAC (Conference held in Sep. 2014), SBMAC.

\bibitem{AAG}
Alonso, R., Amorim, P., Goudon, T.
Analysis of a chemotaxis system arising in ant foraging.
\emph{Submitted.}

\bibitem{Melo2011} Bandeira de Melo, E.B., Araújo, A.F.R.,
Modelling foraging ants in a dynamic and confined environment.
BioSystems 104 (2011) 23--31

\bibitem{Burden}R.L. Burden, J.D. Faires, \emph{Numerical analysis,}
ninth edition, Brooks/Cole (2001).

\bibitem{Beckers1992}
Beckers, R., Deneubourg, J.L., Goss, S., 1992.
Trail laying behaviour during food recruitment in the ant Lasius niger (L.) - Springer
Insectes soc. 39, 1, 59--72.

\bibitem{Bertozzi}
Bertozzi, A.L. \emph{et al.},
Spatiotemporal chemotactic model for ant foraging.
Modern Physics Letters B
Vol. 28, No. 30 (2014) 1450238 

\bibitem{Boissard2012}
Boissard, E., Degond, P., Motsch, S., 2012.
Trail formation based on directed pheromone deposition.
J. Math. Biol., DOI 10.1007/s00285-012-0529-6.

\bibitem{Bossert1963}
Bossert, W.H., Wilson, E.O., The Analysis of Olfactory Communication Among Animals.
J. Theoret. Biol. (1963) 5, 443--469.

\bibitem{Couzin2002}
Couzin, I.D., Franks,. N.R., 2002.
Self-organized lane formation and optimized traffic flow in army ants.
Proc. R. Soc. Lond. B (2003) 270 no.1511 139--146.

\bibitem{Deneubourg1990}
Deneubourg, J.-L., Aron, S., Goss., Pasteels, J.M., 1990.
The Self-Organizing Exploratory Pattern of the Argentine Ant.
Journal of Insect Behavior, Vol. 3, No. 2, 150--168.




\bibitem{Edelstein}
Edelstein-Keshet, L., 1994.
Simple models for trail-following behaviour; Trunk trails versus individual foragers.
J. Math. Biol., 32, 303--328


\bibitem{Edelstein1995}
Edelstein-Keshet, L.,  Watmough, J., and Ermentrout, B.G.
Trail following in ants: individual properties determine population behaviour
Behavioral Ecology and Sociobiology, Vol. 36, No. 2 (1995), pp. 119-133


\bibitem{GR}E. Godlewski, P.-A. Raviart, {\it Hyperbolic systems of conservation laws,} 
Math\'ematiques \& Applications (Paris). Ellipses, Paris, 1991

\bibitem{HillenPainter2001} Hillen, T., and Painter, K.J.
Global Existence for a Parabolic Chemotaxis Model with Prevention of Overcrowding,
Advances in Applied Mathematics 26, 280–301 (2001).

\bibitem{HillenPainter2009}
Hillen, T., and Painter, K.J. A user’s guide to PDE models for chemotaxis, J. Math. Biol. (2009) 58:183–217.

\bibitem{Holldobler1976}
H\"olldobler, B., 1976. Recruitment Behavior, Home Range Orientation and Territoriality in Harvester Ants, Pogonomyrmex.
Behav. Ecol. Sociobiol. 1, 3--44.

\bibitem{TheAnts}
H\"olldobler, B. and Wilson, E.O., 1990. The Ants. The Belknap Press of Harvard University Press, Cambridge, Mass. 

\bibitem{Horstmann1}
Horstmann, D., From 1970 until now: The Keller–Segel model in chemotaxis and its consequences I. Jahresberichte der DMV 105 (2003), 103--165.

\bibitem{Horstmann2}
Horstmann, D. From 1970 until now: The Keller–Segel model in chemotaxis and its consequences II. Jahresberichte der DMV, 106, (2004) pp. 51--69.



\bibitem{Jackson2004}
Jackson, D., Holcombe, M., Ratnieks, F., 2004.
Coupled computational simulation and empirical research into the foraging system of Pharaoh’s ant (Monomorium pharaonis).
Biosystems,
Vol. 76, 1–3, 101--112

\bibitem{Jackson2004-2}
Jackson, D., Holcombe, M., Ratnieks, F., 2004.
Trail geometry gives polarity to ant foraging networks
Nature,
432 (7019), 907--909

\bibitem{Jeanson2003}
Jeanson R., Ratnieks F.L.W., Deneubourg. J.-L., 2003.
Pheromone trail decay rates on different substrates in the Pharaoh’s ant, Monomorium pharaohs. 
Physiological Entomology (2003) 28, 192--198


\bibitem{John}
John, A., \emph{et.~al}, 2009.
Trafficlike collective movement of ants on trails: absence of jammed phase. Phys. Rev. Lett. 102, 108001.

\bibitem{Johnson2006}
Johnson, K., Rossi, L.F., 2006.
A mathematical and experimental study of ant foraging trail dynamics.
Journal of Theoretical Biology, 241, 360–369.   

\bibitem{KellerSegel70}
Keller, E., Segel, L., 1970.
Initiation of slide mold aggregation viewed as an instability. J. Theor. Biol. 26 (1970), 399--415.

\bibitem{KellerSegel71}
Keller, E., Segel, L., 1971. Model for Chemotaxis. J. theor. Biol. 30, 225--234.

\bibitem{LeVeque}R.J. Levee, Finite volume methods for hyperbolic problems. Cambridge Texts in Applied Mathematics. Cambridge University Press, Cambridge (2002).


\bibitem{Lewis93}
Lewis, M.A., J.D. Murray. 1993. Modelling territoriality and wolf-deer interactions. Nature 366:738--740.

\bibitem{Lewis97}
Lewis, M.A., White, K.A.J., and J.D. Murray, 1997.
Analysis of a model for wolf territories.
J. Math. Biol. (1997) 35: 749--774


\bibitem{Motta2011}
Motta Jafelice, R., et al., 2011.
Fuzzy parameters in a partial differential equation model for population dispersal of leaf-cutting ants.
Nonlinear Analysis: Real World Applications 12, 3397--3412.

\bibitem{MoorcroftLewis06}
Moorcroft, P.R, Lewis, M.A., and Robert L Crabtree, R.L. 2006.
Mechanistic home range models capture spatial patterns and dynamics of coyote territories in Yellowstone.
Proc. R. Soc. B 2006 273, 1651--1659.


\bibitem{Muller1988}
M\"uller, M., Wehner, R., 1988.
Path integration in desert ants, Cataglyphis fortis.
Proc. Nat. Acad. Sci. USA, Vol. 85, 5287--5290.

\bibitem{Okubo}
Okubo, A., Levin, S.A. 
\emph{Diffusion and ecological problems: modern perspectives.} Vol. 14 of 
\emph{Interdisciplinary Applied Mathematics.} Berlin: Springer Verlag.

\bibitem{Painter}
Painter, K.J., 2009.
Continuous Models for Cell Migration in Tissues and Applications to Cell Sorting via Differential Chemotaxis.
Bull. Math. Biol. 71, 1117--1147

\bibitem{Patlak}
Patlak, C., 1953. Random walk with persistence and external bias. Bull. Math. Biophys. 15, 311--338.



\bibitem{Ramsch2012}
Ramsch, K., et al., 2012.
A mathematical model of foraging in a dynamic environment by trail-laying Argentine ants.
Journal of Theoretical Biology 306 (2012) 32–45.

\bibitem{Rauch1995}
Rauch, E.M, Millonas, M.M. and Chialvo, D.R. Pattern formation and functionality in swarm models. 
Physics Letters A 207 (1995) 185--193. 

\bibitem{Regnier1968}
Regnier, F.E., Law, J.H. Insect pheromones. J. Lipid Res. 9 (1968) 541--551.

\bibitem{Reid2012}
Reid, C.R., Lattya, T., Beekam, M.
Making a trail: informed Argentine ants lead colony to the best food by U-turning coupled with enhanced pheromone laying. (2012)
Animal Behaviour,
Vol. 84, 6, 1579--1587

\bibitem{Ryan}
Ryan, S.D.
A model for collective dynamics in ant raids. (Under review).


\bibitem{Schweitzer1997}
Schweitzer, F., Lao, K., Family, F. Active random walkers simulate trunk trail formation by ants. BioSystems 41 (1997) 153--166. 

\bibitem{Sumpter2003}
Sumpter, J.T. and Beekman, M. From nonlinearity to optimality: pheromone trail foraging by ants. Animal Behaviour, 2003, 66, 273--280.  

\bibitem{Sumpter2003-2}
Sumpter, D.J.T. and Pratt, S.C.
A modelling framework for understanding social insect foraging. (2003)
Behav Ecol Sociobiol (2003) 53:131--144

\bibitem{Steck2009}
Steck, K., Hansson, B., Knaden, M., 2009. Smells like home- Desert ants, Cataglyphis fortis, use olfactory landmarks to pinpoint the nest.
Frontiers in Zoology, 6:5.

\bibitem{TangTao}
Tang, X., and Tao, Y., 2008. 
Analysis of a Chemotaxis Model for
Multi-Species Host-Parasitoid Interactions
Applied Mathematical Sciences, Vol. 2, no. 25, 1239--1252


\bibitem{FireAnts}
Tschinkel, W.R.,
The Fire Ants,
Harvard University Press, 2006.

\bibitem{Kun2014}
Udiani, O., Pinter-Wollman, N. and Kang, Y.
Identifying robustness in the regulation of collective foraging of ant colonies using an interaction-based model with backward bifurcation.
(Preprint)

\bibitem{Van1986} 
Van Vorhis Key, S.E., Baker, T.C., 1986.
Observations on the Trail Deposition and Recruitment Behaviors of the Argentine Ant, Iridomyrmex humilis (Hymenoptera: Formicidae).
Annals of the Entomological Society of America, 79, 2.  link

\bibitem{Vittori2004}
Vittori, K., et.al., 2004.
Modeling Ant Behavior Under a Variable Environment.
ANTS 2004, LNCS 3172, 190–201. 

\bibitem{Vowles55}
Vowles, D.M., 1955.
The foraging of ants.
The British Journal of Animal Behaviour
3,  1, 1955, 1--13.

\bibitem{Watmough1995}
Watmough, J., Edelstein-Keshet, L., 1995.
A one-dimensional model of trail propagation by army ants.
J. Math. Biology, 33, 459--476 

\bibitem{Watmough1995-2}
Watmough, J., Edelstein-Keshet, L., 1995.
Modelling the Formation of Trail Networks by Foraging Ants.
J. theor. Biol. 176, 357–371

\bibitem{Wehner03}
Wehner, R., 2003.
Desert ant navigation: how miniature brains solve complex tasks.
Journal of Comparative Physiology A, 
189, 8, 579--588

\bibitem{Weyer1985}
Weyer, J., 1985.
A mathematical model for chemical mass recruitment of ants.
J. Math. Biology, 21, 307--315.

\bibitem{Wilson1962-1}
Wilson, E.O., 1962.
Chemical communication among workers of the fire ant Solenopsis saevissima (Fr. Smith) 1. The Organization of Mass-Foraging.
Animal Behaviour,
10, 1-2, 134--138.

\bibitem{Wilson1962-2}
Wilson, E.O., 1962.
Chemical communication among workers of the fire ant Solenopsis saevissima (Fr. Smith) 2. An information analysis of the odour trail.
Animal Behaviour,
10, 1-2, 148--158.

\bibitem{Wilson1962-3}
Wilson, E.O., 1962.
Chemical communication among workers of the fire ant Solenopsis saevissima (Fr. Smith) 3. The experimental induction of social responses.
Animal Behaviour,
10, 1-2, 159--164.


\end{thebibliography}
\end{document}